\documentstyle{l-aa}
\input psfig.sty
\newcommand{\be}{\begin{equation} }
\newcommand{\ee}{\end{equation} }

\begin{document}

\thesaurus{03  % A&A Section 03: Astronomical instrumentation, 
methods and techniques
(11.04.1;  % Galaxies: distances and redshifts
03.13.2 )} % Methods: data analysis

\title{Photometric Redshifts based on standard SED fitting procedures} 

\author{Micol Bolzonella\inst{1,2,3}, Joan-Marc Miralles\inst{4}, 
Roser Pell\'o\inst{3}}

\offprints{M. Bolzonella 
              micol@ifctr.mi.cnr.it, mbolzone@ast.obs-mip.fr}

\institute{Istituto di Fisica Cosmica ``G. Occhialini'', 
via Bassini 15, I-20133 Milano, Italy
\and 
Dipartimento di Fisica, Universit\`a degli Studi di Milano, 
via Celoria 16, I-20133 Milano, Italy
\and
Observatoire Midi-Pyr\'en\'ees, UMR 5572,
14 Avenue E. Belin, F-31400 Toulouse, France
\and
Astronomical Institute, Tohoku University, Aramaki, Aoba-ku, 
Sendai 980-8578, Japan
}

\date{Received, 2000, Accepted , 2000}

\maketitle
\markboth{M. Bolzonella et al.: Photometric $z$ based on standard SED fitting 
procedures}{ }

\begin{abstract}

In this paper we study the accuracy of photometric redshifts computed
through a standard SED fitting procedure, where SEDs are obtained from
broad-band photometry. We present our public code {\it hyperz\/},
which is presently available on the web. We introduce the method and
we discuss the expected influence of the different observational
conditions and theoretical assumptions. In particular, the set of
templates used in the minimization procedure (age, metallicity,
reddening, absorption in the Lyman forest, ...) is studied in detail,
through both real and simulated data. The expected accuracy of
photometric redshifts, as well as the fraction of catastrophic
identifications and wrong detections, is given as a function of the
redshift range, the set of filters considered, and the photometric
accuracy.  Special attention is paid to the results expected from real
data.

\end{abstract}

\section{Introduction}

   The estimate of redshifts through photometry is one of the most
promising techniques in deep universe studies, and certainly a key
point to optimize field surveys with large-field detectors. It is in
fact an old idea of Baum (1962), who originally applied it to the
measure of redshifts for elliptical galaxies in distant clusters. It
was later used by several authors in the eighties (Couch et al. 1983,
Koo 1985) on relatively low-redshift samples, observed in the $\sim
4000$ to $8000$\,\AA\ domain.  Later in the nineties, the interest for
this technique has increased with the development of large field and
deep field surveys, in particular the Hubble Deep Field North and South 
(HDF-N and HDF-S).

Basically two different photometric redshift techniques can be found
in the literature: the so-called empirical training set method, and
the fitting of the observed Spectral Energy Distributions (hereafter 
SED) by synthetic or empirical template spectra.
The first approach, proposed originally by Connolly et al. (1995,
1997), derives an empirical relation between magnitudes and redshifts
using a subsample of objects with measured spectroscopic redshifts,
i.e.  the training set.
A slightly modified version of this method was used by Wang et
al. (1998) to derive redshifts in the HDF-N by means of a linear
function of colours. 
This method produces small dispersions, even when the number of
filters available is small, and it has the advantage that it does not
make any assumption concerning the galaxy spectra or evolution, thus
bypassing the problem of our poor knowledge of high redshift spectra.
However, this approach is not flexible: when different filter sets are
considered, the empirical relation between magnitudes and redshifts
must be recomputed for each survey on a suitable spectroscopic
subsample.
Moreover, the training set is constituted by the brightest objects,
for which it is possible to measure the redshift. Thus, this kind of
procedure could in principle introduce some bias when computing the
redshifts for the faintest sources, because there is no guarantee that
we are dealing with the same type of objects from the
spectrophotometrical point of view.
Also, the redshift range between $1.4$ and $2.2$ had been hardly
reached by spectroscopy up to now, because of the lack of strong
spectral features accessible to optical spectrographs. Thus, no
reliable empirical relation can be found in this interval.

The SED fitting procedure, described in detail in the following
section, bases its efficiency on the fit of the overall shape of
spectra and on the detection of strong spectral properties.
The observed photometric SEDs are compared to those obtained 
from a set of reference spectra, using the same photometric system.
The photometric redshift of a given object corresponds to the best fit
of its photometric SED by the set of template spectra.
This method is used mainly for applications on the HDF, using
either observed or synthetic SEDs (e.g. Mobasher et al. 1996,
Lanzetta et al. 1996, Gwyn \& Hartwick 1996, Sawicki et al. 1997,
Giallongo et al. 1998, Fern\'andez-Soto et al. 1999, Arnouts et
al. 1999, Furusawa et al. 2000). A crucial test in all cases is 
the comparison between the photometric and the spectroscopic redshifts
obtained on a restricted subsample of relatively bright objects.
A combination of this method with the Bayesian marginalization
introducing a prior probability was proposed by Ben\'{\i}tez (2000).

   The aim of this paper is to explain in a straightforward way the
expected performances and limitations of photometric redshifts
computed from broad-band photometry. This study has been conducted
with our public code called {\it hyperz\/}, which adopts a standard
SED fitting method, but most results should be completely general in
this kind of calculations. This program was originally developed by
Miralles (1998) (see also Pell\'o et al. 1999), and the present
version of the code {\it hyperz\/} is available on the web at the
following address: 
\begin{center}
{\tt http://webast.ast.obs-mip.fr/hyperz\/} .
\end{center}

The plan of the paper is the following. In Section~\ref{method} we
present the method used by {\it hyperz\/} and the involved set of
parameters. The accuracy of the redshift determinations and the
expected percentage of catastrophic identifications, as a function of
the filter set and the photometric errors, are studied through
simulations in Section~\ref{simul}.  The influence of the different
parameters on the accuracy of photometric redshifts is investigated in
Section~\ref{param}, using both simulations and spectroscopic data
from the HDF. Section~\ref{realsim} is devoted to the analysis on the
expected accuracy and possible systematics when exploring real data,
coming from deep photometric surveys. A general discussion is given in
Section~\ref{discuss} and conclusions are listed in Section
\ref{conclu}.

\section{The method}
\label{method}

Photometric redshifts (hereafter $z_{\rm phot}$) are based on the
detection of strong spectral features, such as the $4000$\,\AA\ break,
Balmer break, Lyman decrement or strong emission lines. In general,
broad-band filters will allow to detect only ``breaks'', and they are
not sensitive to the presence of emission lines, except when their
contribution to the total flux in a given filter is higher or of the
same order of photometric errors, as it happens in the case of AGNs
(Hatziminaoglou et al. 2000).

\begin{table}[t]
\begin{center}
\begin{tabular}{ccc}
\hline
Filter       &  $\lambda_{\rm eff}$\,[\AA]  &  width\,[\AA]  \\
\hline \hline
$U$     &   3652  &   543 \\
$B$     &   4358  &   987 \\
$V$     &   5571  &  1116 \\
$R$     &   6412  &  1726 \\
$I$     &   7906  &  1322 \\
$Z$     &   9054  &  1169 \\
$J$     &  12370  &  2034 \\
$H$     &  16464  &  2863 \\
$K$     &  22105  &  3705 \\
F300W   &   3010  &   854 \\
F450W   &   4575  &   878 \\
F606W   &   6039  &  1882 \\
F814W   &   8010  &  1451 \\
\hline 
\end{tabular}
\caption{Characteristics of filters used in the simulations: the 
effective wavelength $\lambda_{\rm eff}$ and the surface of the 
normalized response function.}
\label{tabfilt}
\end{center}
\end{table}

The method used in this paper to compute photometric redshifts is a
SED fitting through a standard $\chi^2$ minimization procedure,
computed with our code {\it hyperz\/}. The observed SED of a given
galaxy is compared to a set of template spectra:

\begin{equation}
 \chi^2(z)=\sum_{i=1}^{N_{\rm filters}} \left[{F_{{\rm obs},i}
- b \times F_{{\rm temp},i}(z)\over \sigma_i} \right]^2 \:,
\label{eq1}
\end{equation}
where $F_{{\rm obs},i}$, $F_{{\rm temp},i}$ and $\sigma_i$ are the
observed and template fluxes and their uncertainty in filter $i$,
respectively, and $b$ is a normalization constant. 

   The new Bruzual \& Charlot evolutionary code (GISSEL98, Bruzual \&
Charlot 1993) has been used to build 8 different synthetic
star-formation histories, roughly matching the observed properties of
local field galaxies from E to Im type: a delta burst, a constant
star-forming system, and six $\mu$-models (exponentially decaying SFR)
with characteristic time-decays chosen to match the sequence of
colours from E-S0 to Sd. We use the Initial Mass Function (IMF) by
Miller \& Scalo (1979), but this choice has a negligible impact on
the final results, as discussed in Section~\ref{imf}. 
The upper mass limit for star formation is $125\,M_{\odot}$.  
The basic database includes only solar metallicity SEDs,
but other possibilities are discussed in Section~\ref{param}. The
library also includes a set of empirical SEDs compiled by Coleman, Wu
and Weedman (1980) (hereafter CWW) to represent the local population
of galaxies. CWW spectra were extended to wavelengths $\lambda \le
1400$\,\AA\ and $\lambda \ge 10000$\,\AA\ using the equivalent GISSEL
spectra.  The synthetic database derived from Bruzual \& Charlot
includes 408 spectra (51 different ages for the stellar population and
8 star-formation regimes).  In most applications, there is no sensible
gain when the number of $\mu$-models is reduced to only 3, thus
including only 255 spectra.

   Throughout this paper we use the same set of broad-band filters,
with characteristics presented in Table~\ref{tabfilt}. These filters
cover all the wavelength domain under study, without major overlap or
gap. We also include the HDF filters used in Sections~\ref{param} and
\ref {realsim} (from Biretta et al. 1996).  The {\it hyperz\/} filter
library is an enlarged version of the original Bruzual \& Charlot one,
and presently includes 163 filters and detector responses. All
magnitudes given in this paper refer to the Vega system.

   {\it Hyperz\/} has been optimized to gain in efficiency when
computing $z_{\rm phot}$ on large catalogues. The input data for a
given catalogue are magnitudes and photometric errors. To compute a 
reliable estimate of $z_{\rm phot}$, the colours and the corresponding 
photometric errors must be obtained with particular care, including 
uncertainties due to zero-points, intrinsic accuracy, etc. Magnitudes 
are obtained within the same aperture in all filters, after correction 
for seeing differences between images. 
For a given catalogue, the relevant parameters introduced in the
$z_{\rm phot}$ calculation are:

\begin{itemize}

\item The set of template spectra. This point includes the SFR type,
the possible link between the age and the metallicity of the stellar
population, and the choice of an IMF. It is discussed in Section
\ref{param}. 

\item The reddening law is usually taken from Calzetti et
al. (2000), but 4 other laws are also included in the code. This is 
discussed in Section~\ref{red}.
The input value is $A_V$, corresponding to a dust-screen model, with 
$F_{\rm o}(\lambda)=F_{\rm i}(\lambda) 10^{-0.4 A_{\lambda}}$,
where $F_{\rm o}$ and $F_{\rm i}$ are the observed and the intrinsic 
fluxes, respectively. 
The extinction at a wavelength $\lambda$ is related to the colour 
excess $E_{B-V}$ and to the reddening curve $k(\lambda)$ by
$A_{\lambda} = k(\lambda) E_{B-V} = k(\lambda) A_V/R$, with $R=3.1$
except for the Small Magellanic Cloud ($R=2.72$) and the Calzetti's
law ($R=4.05$).
The normal setting for $A_V$ ranges between 0 and $1.5$ magnitudes.
The mean galactic extinction correction towards a given line of sight
can be introduced in terms of 
$E_{B-V}$, and it is applied to the whole catalogue.

\item Flux decrements in the Lyman forest are computed according to
Giallongo \& Cristiani (1990) and Madau (1995), both of them giving
similar results. 

\item The limiting magnitude in each filter, and the rule to be
applied in the case of non detection. The rule is set for each filter
independently, and there are 4 different possibilities:
0) the filter is not taken into account in the computation;
1) the flux in this filter is set to 0 with an error bar corresponding
to the flux deduced from the limiting magnitude;
2) the flux in this filter is set to $1/2$ of the limiting flux,
according to the limiting magnitude, and the associated 1 sigma error
is $\pm 1/2$ times this value; 
3) the flux and the 1 sigma error in this filter are computed from the
limiting magnitude and from the error associated to the limiting
magnitude (both fixed). 
Case 1 is the usual setting when one is dealing with a relatively 
deep survey in the considered  filter, whereas case 0 applies to 
``out-of-field'' objects. Case 2 and 3 are well suited for relatively 
shallow surveys. 
The idea of ``shallow'' and ``deep'' in this context refers to relative 
values of the limiting magnitudes associated to the different filters
in the photometric catalogue.

\item The cosmological parameters $H_0$, $\Omega_0$ and
$\Omega_{\Lambda}$, which are only related here to the maximum age
allowed to the stellar population at a given redshift. The age
checking is an option.

\end{itemize}

\begin{figure}
\centering\psfig{file=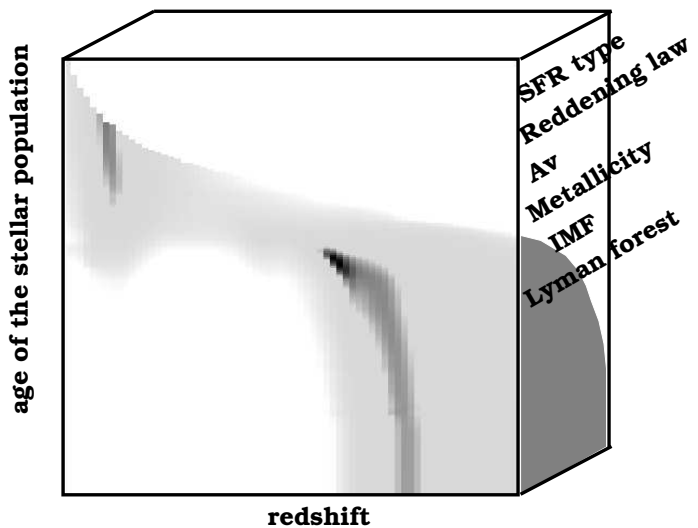,angle=270,width=0.49\textwidth}
\caption{Artist view of the SED fitting procedure to compute $z_{\rm
phot}$. The figure presents a likelihood map for a
representative object at $z \sim 4$. The shaded area encloses the
highest confidence level region according to the $\chi^2$ associated 
probability.
Each point on the redshift-age map 
corresponds to the best fit of the SED obtained 
across the parameter space. 
The degeneracy in the parameter space is shown in this example. }
\label{hyper}
\end{figure}

   Due to the degeneracy in the parameter space defined by the SFR
type, age, metallicity and reddening, the $z_{\rm phot}$ computation
for a given object is equivalent to finding the most likely solution
for the redshift across this parameter space, regardless to details on
the best-fit SED (see Figure~\ref{hyper}).
Both the $z_{\rm phot}$ and the SED are obtained through {\it hyperz\/},
together with the best fit parameters ($A_V$, spectral type,
metallicity and age). Because of the degeneracy between these
parameters, the relevant information shall be the redshift and the
rough SED type, in the sense that a given object has a ``blue" or
``red" continuum at a given $z$, but no reliable information can be
obtained about the other parameters from broad-band photometry alone.
 
\section{Filters and photometric accuracy}
\label{simul}

\begin{figure*}
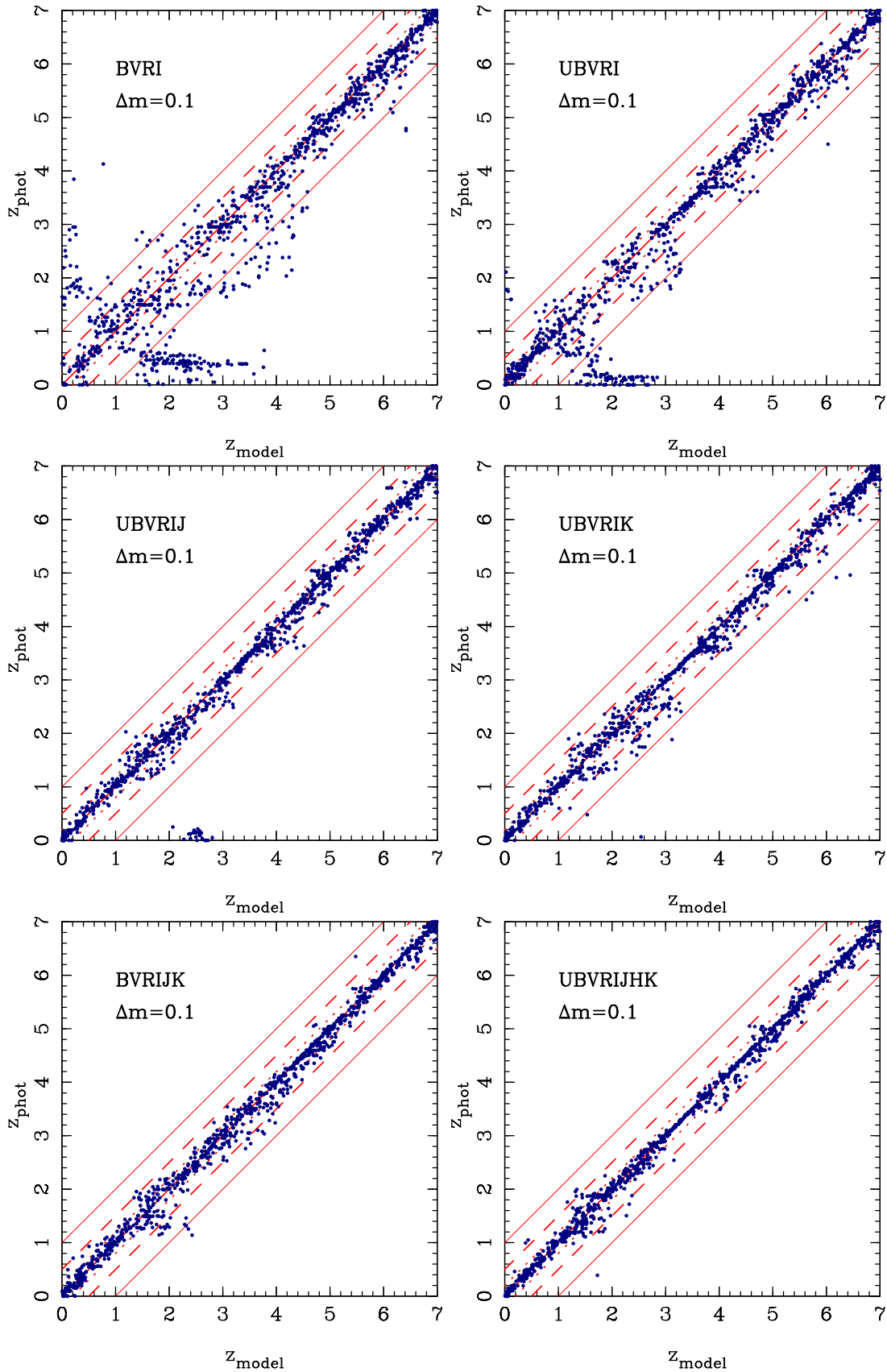

{\centering \leavevmode
\psfig{file=9769_f2a.ps,width=.40\textwidth}
\psfig{file=9769_f2b.ps,width=.40\textwidth}
\newline
\psfig{file=9769_f2c.ps,width=.40\textwidth}
\psfig{file=9769_f2d.ps,width=.40\textwidth}
\newline
\psfig{file=9769_f2e.ps,width=.40\textwidth}
\psfig{file=9769_f2f.ps,width=.40\textwidth} }
\caption{Comparison between $z_{\rm model}$ and $z_{\rm phot}$ for 
simulated catalogues with $\Delta m =0.1$ and filters sets {\em BVRI}, 
{\em UBVRI}, {\em UBVRIJ}, {\em UBVRIK}, {\em BVRIJK}, {\em UBVRIJHK}. 
Dotted lines correspond to $\Delta_z=0.2$, 
dashed lines to $\Delta_z=0.5$ and thin solid lines to $\Delta_z=1$. }
\label{figsim}
\end{figure*}

\tabcolsep 0.2cm
\begin{table*}
\begin{center}
\begin{tabular}{|c|c|cc|cc|cc|cc|cc|cc|}
\hline
 & & 
\multicolumn{2}{c|}{$z = 0.0$ -- 0.4} & \multicolumn{2}{c|}{0.4 -- 1.0} & 
\multicolumn{2}{c|}{1.0 -- 2.0} & \multicolumn{2}{c|}{2.0 -- 3.0} & 
\multicolumn{2}{c|}{3.0 -- 5.0} & \multicolumn{2}{c|}{5.0 -- 7.0} \\
\cline{3-14}
filters &$\Delta m$& 
$\sigma_z$ & $\left<\Delta_z\right>$  & 
$\sigma_z$ & $\left<\Delta_z\right>$  & 
$\sigma_z$ & $\left<\Delta_z\right>$  & 
$\sigma_z$ & $\left<\Delta_z\right>$  &
$\sigma_z$ & $\left<\Delta_z\right>$  & 
$\sigma_z$ & $\left<\Delta_z\right>$  \\
\hline \hline
               & 0.05 & 0.18 &  0.00   
                      & 0.22 & -0.13   
                      & 0.32 &  0.10   
                      & 0.30 &  0.07  
                      & 0.23 &  0.08   
                      & 0.16 &  0.05  \\ 
{\em BVRI}     & 0.10 & 0.20 & -0.05   
                      & 0.30 & -0.21 
                      & 0.38 &  0.08  
                      & 0.45 &  0.10  
                      & 0.25 &  0.10 
                      & 0.19 &  0.07  \\
               & 0.20 & 0.29 & -0.19  
                      & 0.37 & -0.24  
                      & 0.41 &  0.18 
                      & 0.54 &  0.01 
                      & 0.34 &  0.14 
                      & 0.23 &  0.07  \\
               & 0.30 & 0.25 & -0.34 
                      & 0.38 & -0.24  
                      & 0.42 &  0.25  
                      & 0.53 & -0.21 
                      & 0.35 &  0.14  
                      & 0.28 &  0.09  \\
\hline
               & 0.05 & 0.07 & -0.03  
                      & 0.17 & -0.07 
                      & 0.26 &  0.12  
                      & 0.21 &  0.04 
                      & 0.17 &  0.05  
                      & 0.18 &  0.06  \\
{\em UBVRI}    & 0.10 & 0.09 & -0.03  
                      & 0.21 & -0.11  
                      & 0.35 &  0.17  
                      & 0.33 &  0.11 
                      & 0.23 &  0.08 
                      & 0.19 &  0.04  \\
               & 0.20 & 0.20 & -0.11  
                      & 0.29 & -0.19  
                      & 0.42 &  0.17 
                      & 0.41 &  0.12 
                      & 0.27 &  0.09  
                      & 0.23 &  0.06  \\
               & 0.30 & 0.28 & -0.20  
                      & 0.31 & -0.18 
                      & 0.49 &  0.16  
                      & 0.45 &  0.11  
                      & 0.29 &  0.11 
                      & 0.27 &  0.07  \\ 
\hline
               & 0.05 & 0.04 & -0.01  
                      & 0.11 & -0.05  
                      & 0.25 &  0.11  
                      & 0.13 &  0.04  
                      & 0.17 &  0.06  
                      & 0.09 &  0.01  \\
{\em UBVRIZ}   & 0.10 & 0.07 & -0.02  
                      & 0.16 & -0.08  
                      & 0.28 &  0.11  
                      & 0.23 &  0.07  
                      & 0.22 &  0.09 
                      & 0.14 &  0.02  \\
               & 0.20 & 0.17 & -0.08  
                      & 0.22 & -0.11  
                      & 0.41 &  0.11  
                      & 0.34 &  0.16  
                      & 0.28 &  0.12  
                      & 0.19 &  0.05  \\
               & 0.30 & 0.21 & -0.12  
                      & 0.27 & -0.15  
                      & 0.44 &  0.12 
                      & 0.40 &  0.18 
                      & 0.32 &  0.15  
                      & 0.24 &  0.08  \\
\hline
               & 0.05 & 0.04 & -0.01  
                      & 0.07 & -0.02  
                      & 0.11 & -0.01  
                      & 0.12 &  0.07  
                      & 0.12 &  0.04  
                      & 0.11 &  0.02  \\
{\em UBVRIJ}   & 0.10 & 0.08 & -0.01  
                      & 0.11 & -0.06  
                      & 0.20 & -0.01  
                      & 0.14 &  0.10  
                      & 0.17 &  0.07 
                      & 0.15 &  0.03  \\
               & 0.20 & 0.17 & -0.09  
                      & 0.19 & -0.08  
                      & 0.30 &  0.00  
                      & 0.23 &  0.14  
                      & 0.26 &  0.13  
                      & 0.18 &  0.03  \\
               & 0.30 & 0.22 & -0.13  
                      & 0.26 & -0.15  
                      & 0.35 &  0.01 
                      & 0.30 &  0.17 
                      & 0.30 &  0.17  
                      & 0.22 &  0.06  \\
\hline
               & 0.05 & 0.04 & -0.01  
                      & 0.07 & -0.02  
                      & 0.11 &  0.00  
                      & 0.21 &  0.07  
                      & 0.13 &  0.04 
                      & 0.15 &  0.05  \\
{\em UBVRIK}   & 0.10 & 0.08 & -0.02  
                      & 0.10 & -0.05  
                      & 0.22 &  0.01  
                      & 0.27 &  0.12  
                      & 0.18 &  0.08  
                      & 0.15 &  0.06  \\
               & 0.20 & 0.17 & -0.064 
                      & 0.16 & -0.08  
                      & 0.31 &  0.01  
                      & 0.32 &  0.18  
                      & 0.22 &  0.12  
                      & 0.21 &  0.08  \\
               & 0.30 & 0.22 & -0.14  
                      & 0.25 & -0.14  
                      & 0.36 & -0.01  
                      & 0.35 &  0.20 
                      & 0.26 &  0.14  
                      & 0.23 &  0.09  \\
\hline
               & 0.05 & 0.06 & -0.01  
                      & 0.06 & -0.03  
                      & 0.16 &  0.00  
                      & 0.16 &  0.01  
                      & 0.13 &  0.03  
                      & 0.09 &  0.03  \\
{\em BVRIJK}   & 0.10 & 0.12 & -0.03  
                      & 0.11 & -0.05  
                      & 0.24 &  0.00  
                      & 0.19 &  0.02  
                      & 0.18 &  0.07  
                      & 0.12 &  0.04  \\
               & 0.20 & 0.19 & -0.06  
                      & 0.23 & -0.07  
                      & 0.33 & -0.04  
                      & 0.27 &  0.06  
                      & 0.23 &  0.11  
                      & 0.16 &  0.07  \\
               & 0.30 & 0.25 & -0.14  
                      & 0.26 & -0.14  
                      & 0.40 & -0.04  
                      & 0.32 &  0.09 
                      & 0.27 &  0.16  
                      & 0.20 &  0.09  \\
\hline
               & 0.05 & 0.04 & -0.01  
                      & 0.05 & -0.02  
                      & 0.13 &  0.02  
                      & 0.06 &  0.00  
                      & 0.12 &  0.03  
                      & 0.09 &  0.02  \\
{\em UBVRIJK}  & 0.10 & 0.09 & -0.01  
                      & 0.09 & -0.03  
                      & 0.21 &  0.03  
                      & 0.11 &  0.02  
                      & 0.15 &  0.05  
                      & 0.12 &  0.03  \\
               & 0.20 & 0.18 & -0.05  
                      & 0.18 & -0.06  
                      & 0.32 &  0.03  
                      & 0.20 &  0.05  
                      & 0.18 &  0.09  
                      & 0.15 &  0.05  \\
               & 0.30 & 0.23 & -0.09  
                      & 0.24 & -0.10  
                      & 0.33 &  0.05  
                      & 0.28 &  0.08  
                      & 0.21 &  0.11  
                      & 0.19 &  0.07  \\
\hline
               & 0.05 & 0.03 &  0.00  
                      & 0.05 & -0.01  
                      & 0.10 &  0.00  
                      & 0.06 &  0.00  
                      & 0.09 &  0.01  
                      & 0.09 &  0.02  \\
{\em UBVRIJHK} & 0.10 & 0.10 & -0.02  
                      & 0.10 & -0.02  
                      & 0.20 & -0.01  
                      & 0.13 &  0.03  
                      & 0.13 &  0.04  
                      & 0.11 &  0.03  \\
               & 0.20 & 0.18 & -0.06  
                      & 0.17 & -0.07  
                      & 0.28 & -0.02  
                      & 0.19 &  0.06  
                      & 0.19 &  0.09  
                      & 0.16 &  0.06  \\
               & 0.30 & 0.25 & -0.12  
                      & 0.25 & -0.11  
                      & 0.33 & -0.04  
                      & 0.27 &  0.09  
                      & 0.23 &  0.11  
                      & 0.21 &  0.09  \\
\hline
\end{tabular}
\caption{Summary of results obtained on simulated catalogues with a
homogeneous redshift distribution as a function of the redshift bin,
filters set and photometric errors $\Delta m$. See the text for a
complete description.}
\label{tabsim}
\end{center}
\end{table*}

   In this section we study through simulations the quality of the
$z_{\rm phot}$ as a function of the filter set, the photometric
accuracy and the redshift, i.e. the robustness of the redshift
determination and the expected percentage of catastrophic
identifications and spurious detections.  The aim of this exercise is
to study the systematic effects produced by the sampling of the SED
and the associated noise coming from photometry. 
Catastrophic identifications ($l\%$) are those with $ | \Delta_z | = |
z_{\rm model} - z_{\rm phot} | \ge 1$, and such objects are thus lost
from their original redshift bin.
The accuracy of $z_{\rm phot}$ in a given redshift bin is defined by
the mean difference $\left<\Delta_z \right>=\sum \Delta_z /N$ of the
sample with respect to the model redshift, excluding catastrophic
identifications, and the standard deviation $\sigma_z = \sqrt{\sum
(\Delta_z-\left<\Delta_z \right>)^2/(N-1)}$.
Spurious identifications ($g\%$) correspond to objects which are
incorrectly assigned to a given $z_{\rm phot}$ interval, and thus
susceptible to contaminate the statistics within this $z_{\rm phot}$
interval; in this case $ | \Delta_z | \ge 3 \times \sigma_z$.

Some of these quantities, in particular $l\%$ and $g$\%, depend on
assumptions about redshift number counts and photometric depth.  For
this reason we compute them only for a set of simulations with a more
realistic modeling for galaxy counts, according to a Pure Luminosity
Evolution (PLE) scenario. We discuss the results as a function of the
photometric parameters in Section~\ref{realsim}.

   Simulated catalogues of $1000$ objects were produced, with a
homogeneous redshift distribution, in order to compute the above
mentioned parameters as a function of the filter set and photometric
accuracy.  In all cases, the types and ages assigned to the different
galaxies in a redshift bin are randomly chosen from the 8 GISSEL98
template families mentioned above, with solar metallicity.
Photometric errors in these homogeneous catalogues are introduced as a
noise following a Gaussian distribution of fixed $1\sigma$ in
magnitudes for each band (0.05 to 0.3 magnitudes, i.e. $\sim$ 5 to 30
\% photometric accuracy), and they are uncorrelated for different
filters. For each filter set we study the quality of $z_{\rm phot}$ as
a function of the photometric accuracy, In this particular case,
photometric errors do not scale with magnitudes.  A realistic error
distribution is used in Section~\ref{realsim}.  The value of the
visual extinction $A_V$ ranges between $0$ and $1$.  For each
simulated galaxy, {\it hyperz\/} computes a $z_{\rm phot}$ value, as
well as the $z_{\rm phot}$ error bars corresponding to $P=68, 90,
99$\% confidence levels, computed by means of the $\Delta \chi^2$
increment for a single parameter (Avni 1976). The redshift step used
to search solutions between $z=0$ and $z=7$ is $\Delta z = 0.05$, with
an internal accuracy which is 10 times better. The choice of the
primary $z$-step between 0.1 and 0.05 does not affect significantly
the results.

\begin{figure*}
{\centering \leavevmode
\psfig{file=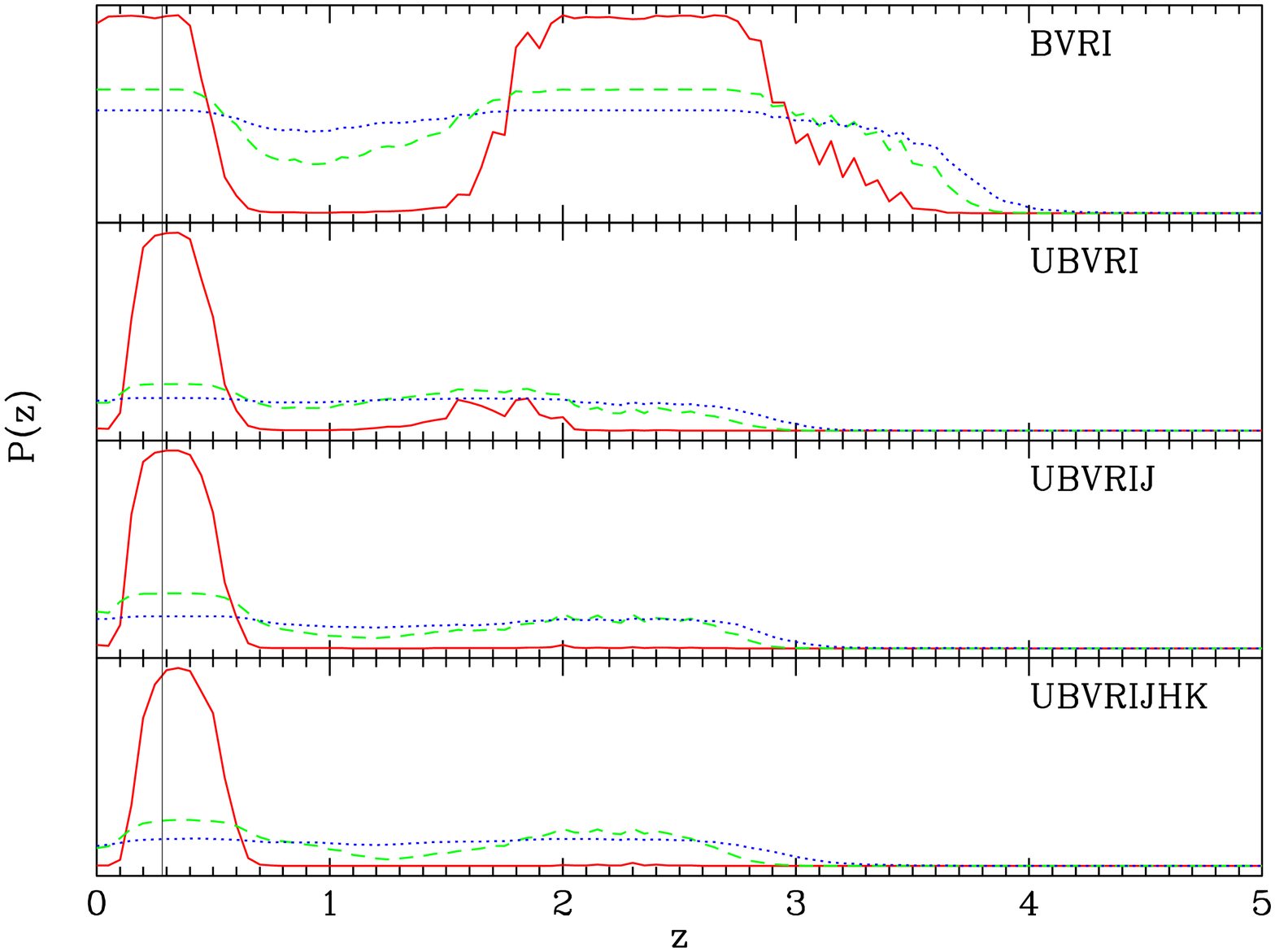,width=.49\textwidth,angle=270}
\psfig{file=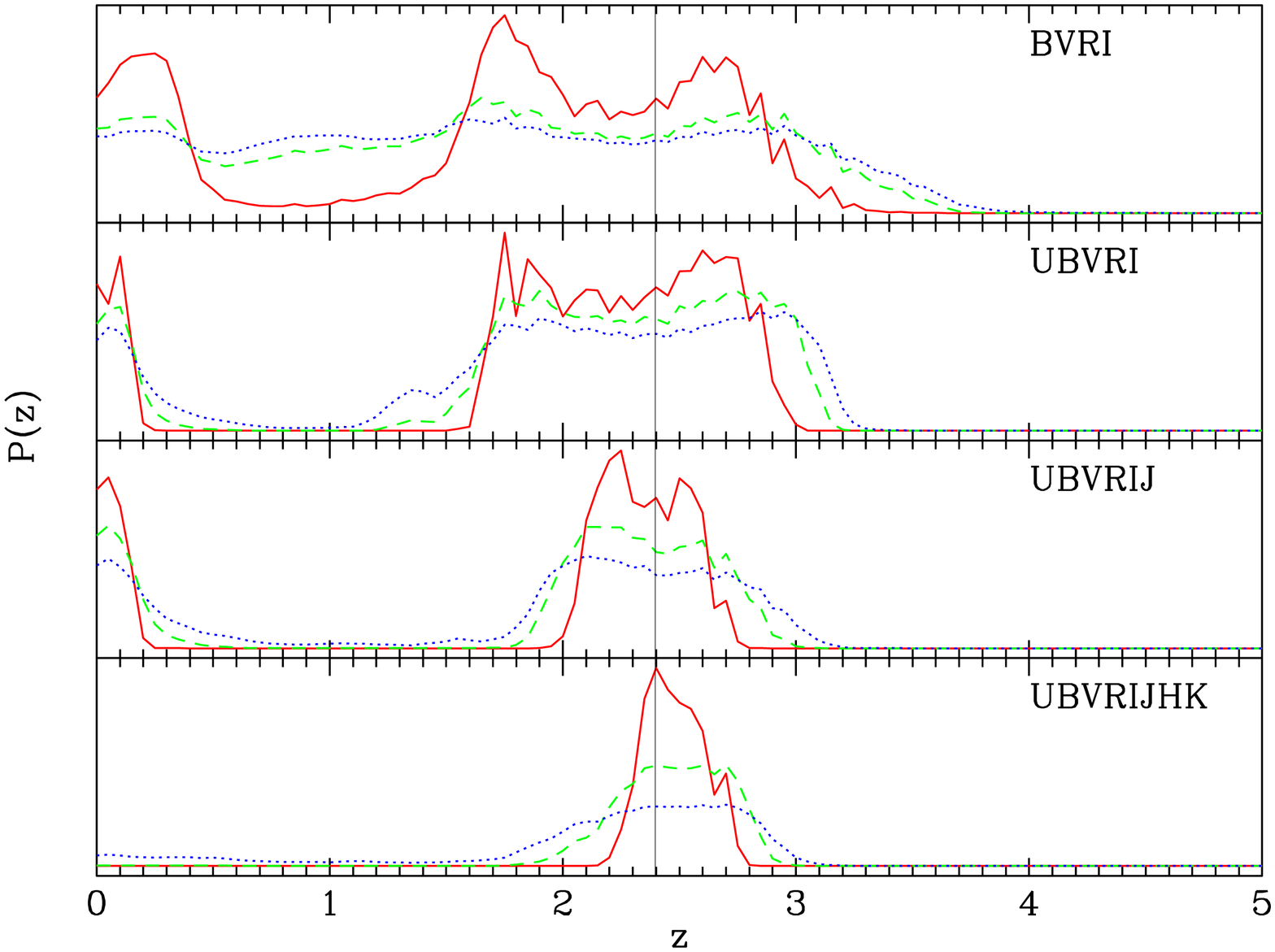,width=.49\textwidth,angle=270}}
\caption{Examples of the evolution of the probability distributions 
associated to  $\chi^2(z)$ as a function of the filter set and 
photometric errors, for two simulated objects. 
Left:  $z_{\rm model}=0.282$.
Right: $z_{\rm model}=2.396$. 
Dotted lines refer to $\Delta m=0.3$, dashed lines: $\Delta m=0.2$, 
solid lines: $\Delta m=0.1$. The vertical line marks the true
$z_{\rm model}$ value.}
\label{like}
\end{figure*}

   Figure~\ref{figsim} shows the behaviour of the different sets of
simulated samples when the $z_{\rm phot}$ is compared to the true
$z_{\rm model}$. The results of these simulations are summarized in
Table~\ref{tabsim}. Without near-IR photometry, the errors on
individual galaxies become huge at $1.2 \le z \le 2.2$ as expected due
to the lack of strong spectral features in the visible band.  In
particular, in this redshift range the 4000\,\AA\ break goes out of
the $I$ band and the Lyman break does not yet affect the photometry in
the filter $U$.
This problem is solved when near-IR is included. In fact, $J$, $H$,
and $K$ filters allow to bracket the 4000\,\AA\ break.
Also, the lack of $U$ band photometry introduces an enhanced
uncertainty in the $z \le 0.4$ domain (mainly because of the
contribution at $z \la 0.2$), because at $z \le 0.2$ none of the
other filters is able to detect a strong break.

All these results are almost independent on the type of galaxy,
provided that the evolving population of stars is older than a few
$\sim 10^{7}$ years typically. This point is discussed in details in
next section.

    The dispersion in $z_{\rm phot}$ is strongly sensitive to the
photometric uncertainties.  There is no significant gain for $\Delta m
\le 0.05$ magnitudes (about $5$\% accuracy). This value roughly
corresponds to the typical photometric uncertainties in deep
photometric surveys, when all the error sources are included.  The
dispersion and the number of multiple solutions with similar weight
rapidly increase up to $\Delta m \sim 0.3$ magnitudes.  Including
near-IR {\em JHK\/} photometry strongly reduces the error bars within
the $1.2 \le z \le 2.2$ range, without significantly improving the
uncertainties in $z_{\rm phot}$ outside this interval.  If the filter
$Z$ is considered in addition to the five optical filters, the
resulting dispersion at low redshift become smaller up to $z_{\rm
model} \simeq 1.5$, but the degeneracy at $z_{\rm model}=1.5$ -- $3$
still remains, even if less dramatic.

In Figure~\ref{like} we illustrate the probability functions for two
simulated galaxies at low and high redshift: the solution becomes
better constrained around the model value and the degeneracy between
high and low redshift solutions disappears with increasing photometric
accuracy and when the wavelength range extends up to the near infrared
region.

   The typical dispersion in $z_{\rm phot}$ obtained here is similar to
the values found in the literature, even when the techniques used are
appreciably different (Brunner et al. 1997, Connolly et al 1997,
...). In most published studies it is extremely difficult to compare
the accuracy of $z_{\rm phot}$ as a function of photometric errors.

These results are useful to understand the general trends expected 
from a given configuration of filters and photometric accuracy. 
Nevertheless, $z_{\rm phot}$ techniques are often applied to statistical 
studies, which require more ``realistic" simulations in order to 
define the right observational strategy for the photometric survey. 
Then, a realistic redshift distribution is needed.
For most applications, a PLE model is enough to determine the
main trends. Also, photometric uncertainties have to be scaled with
magnitude, to reproduce the behaviour of real catalogues. 
These points are discussed in Section~\ref{realsim}.

\section{Influence of the different parameters on $z_{\rm phot}$ accuracy}
\label{param}

\subsection{Templates and Lyman forest blanketing}

   We have studied the influence of the set of templates used on the
final results through a comparison between the {\it hyperz\/} $z_{\rm
phot}$ determinations and real spectroscopic data on HDF. All the
other parameters are fixed in this case, and the only difference is
the set of templates used to compute $z_{\rm phot}$.
Table~\ref{tabdelta} summarizes these results.
A similar blind test was recently performed by Hogg et al. (1998) on a
sample of HDF-N galaxies at $z<1.4$, using different procedures and, in
particular, different sets of templates.

We have computed photometric redshifts for the sample of 108 galaxies
on the HDF-N with observed $z_{\rm spec}$ (Cohen et al. 1996; Cowie
1997; Zepf et al. 1996; Steidel et al. 1996; Lowenthal et al. 1997)
considered by Fern\'andez-Soto et al. (1999) plus 4 galaxies from
HDF-S (Glazebrook et al. 2000, in preparation).
Among these, 83 galaxies are at $z<1.5$ and 29 at $2<z<6$.  Photometry
was obtained from the Stony Brook's group (Fern\'andez-Soto et
al. 1999, SUNY web pages {\tt http://www.ess.sunysb.edu/astro/hdfs}) 
using the package SExtractor (Bertin \& Arnouts 1996) to detect sources,
and consists in 7 filters for the HDF-N (F300W, F450W, F606W, F814W
plus near infrared photometry in {\em JHK\/} filters obtained by
Dickinson et al. 1998 at the KPNO IRIM camera) and 12 for the HDF-S
(F300W, F450W, F606W, F814W, plus an additional shallow optical
catalogue {\em UBVRI\/} from NTT SUSI2, and near infrared {\em JHK\/}
data  obtained with NTT SOFI). 
Here we consider results obtained using the 7 filters for the HDF-N
galaxies and all the 12 available filters for the four objects of the
HDF-S subsample. Calculations on the HDF-S using 7 filters do not
affect significantly the individual photometric redshift and the
overall statistic.

To calculate magnitudes from the available measured fluxes in the
catalogues, we considered as non-detection criterion a signal-to-noise
ratio $S/N<1$. In this case we assigned a magnitude $=99$ and we used
the information about the limiting magnitude in the involved filter.

Three different sets of templates are considered in this section: the
basic 5 GISSEL98 models with solar metallicity mentioned above (1
delta burst, 3 $\mu$-decaying, 1 constant star-formation system), the
CWW set of empirical SEDs, and the CWW set extended with a SED of a
very blue galaxy taken from GISSEL library (Miller \& Scalo IMF,
constant SFR, ${\rm age}=0.1$\,Gyr). Adding new very blue spectra to
the third set does not change perceptibly the results.
As for the simulated catalogues, we search solutions in the redshift
interval $z=0$ -- $7$ with a step $\Delta z=0.05$.
In all cases, a crude limit in absolute magnitude has been imposed to
compute $z_{\rm phot}$, with $M_B \in [-28,-9]$.
Moreover, we checked the age of the template to be consistent with the
age of the universe at the considered redshift, depending on the
cosmological model. Here we use $\Omega_0=1$, $\Omega_\Lambda = 0$ and
$H_0=50\,{\rm km \,s^{-1} \,Mpc^{-1}}$.  The reddening is assumed to
range from $A_V=0$ to $1.2$, following the Calzetti et al. (2000) law.

\begin{figure*}
\centering\psfig{file=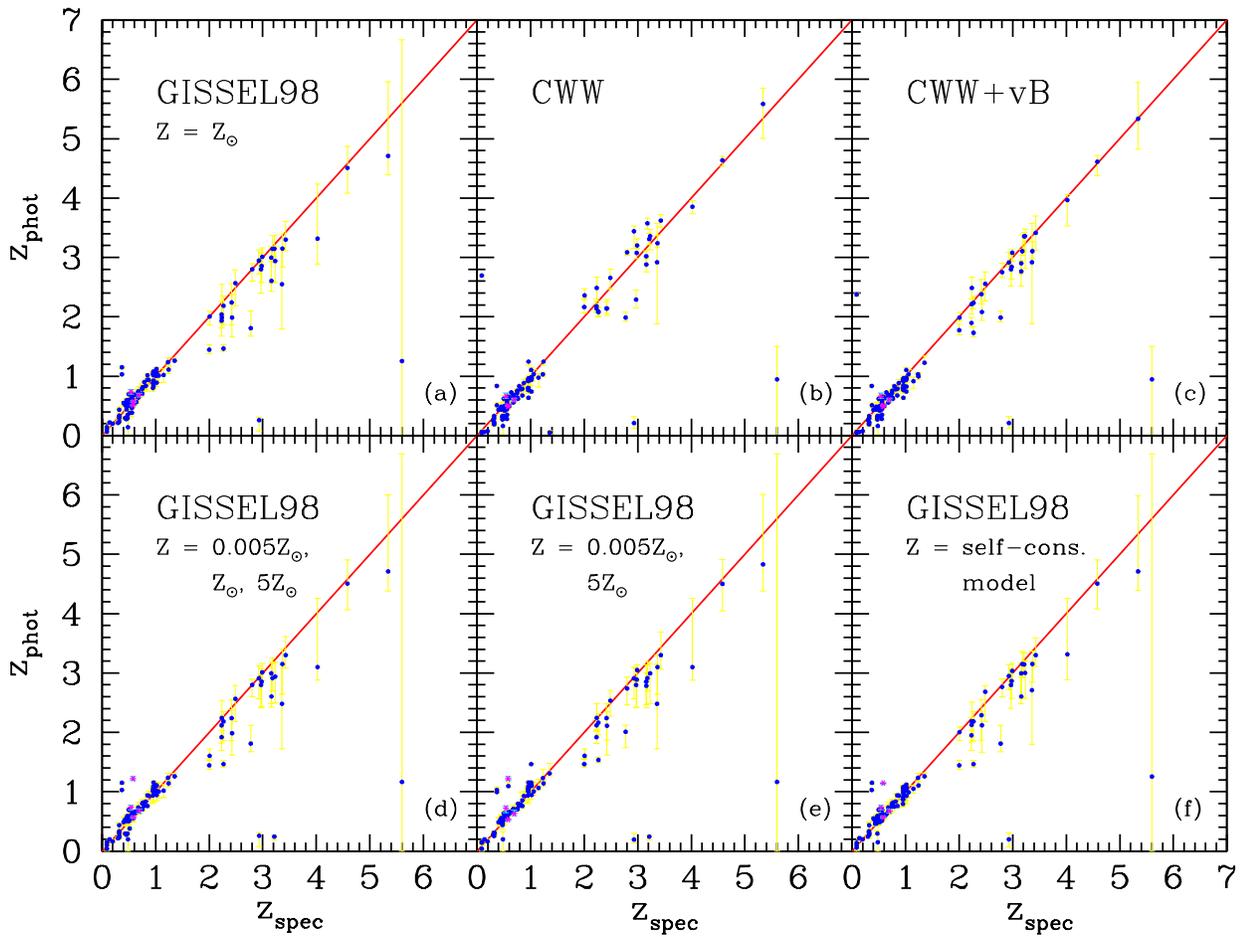,width=0.99\textwidth,angle=270}
\caption{Comparison between photometric and spectroscopic redshifts
for the HDF spectroscopic sample. Error bars in $z_{\rm phot}$
correspond to $3\sigma$. Different sets of template spectra are used
in the various panels: (a) the basic 5 GISSEL98 models with solar
metallicity; (b) the CWW set of empirical SEDs; (c)the CWW set
extended with a SED of a very blue galaxy. The lower panels (d), (e)
and (f) present the comparison between template sets of different
metallicities. See text for more details.}
\label{hdf}
\end{figure*}

The comparison between $z_{\rm spec}$ and $z_{\rm phot}$ for the 112
galaxies of the sample is shown in Figure~\ref{hdf}, for the three
sets of templates hereafter referenced as (a), (b) and (c),
respectively.  Each of them produces a fairly good agreement with the
measured spectroscopic redshifts, but noticeable differences appear
when considering the values of the dispersion, computed as

\[\delta_z = \frac{\sigma}{1+\left< z \right>}
= \sqrt{\frac{\sum_{i=1}^{N}|z_{{\rm phot},i}-z_{{\rm spec},i}|^2}{N-1}}
\frac{1}{1+\left< z \right>}\:,\] 
in the two redshift domains ($z<1.5$ and $2<z<6$):

\begin{itemize} 

\item for $z_{\rm spec}<1.5$ ($\left< z \right> \simeq 0.65$) we
found: (a) $\delta_z = 0.09$; (b) 0.21; (c) 0.17.
If we exclude objects with $|z_{\rm phot}-z_{\rm spec}|>0.5$, the
value of $\delta_z$ reduces to $0.06$ in case (a) (81 objects), and to
$0.08$ for (b) and (c) (81 and 82 objects respectively).  In case (a),
these rejected objects correspond to the two galaxies with uncertain
spectroscopic redshifts (see Arnouts et al. 1999).
Thus, using GISSEL models in this redshift domain produces more
accurate results than CWW templates alone.

\item in the high redshift domain, the dispersion in all the
considered cases is $\delta_z = 0.26$ ($\left< z \right> \simeq
3.06$). If we remove catastrophic identifications (2 objects),
characterized by $|z_{\rm phot}-z_{\rm spec}|>1$, then $\delta_z =
0.10$ in case (a), $0.08$ in (b) and $0.07$ in (c).  In this
case, the CWW set produces slightly better results than GISSEL,
probably due to a Lyman blanketing effect. This point is discussed
below.

\end{itemize}

\begin{table*}[t]
\begin{center}
\begin{tabular}{lccccc}
\hline
             &   &\multicolumn{2}{c}{$z_{\rm spec}<1.5$}
                 &\multicolumn{2}{c}{$2<z_{\rm spec}<6$} \\
             &   & $\delta_z$         & $\delta_z$ 
                 & $\delta_z$         & $\delta_z$  \\
Template set &   & (all, $N=83$)      & (non-cata.)
                 & (all, $N=29$)      & (non-cata.) \\
\hline \hline
GISSEL ($Z_{\odot}$) 
             &(a)&  0.09   &  0.06(81)    &  0.26   &  0.10(27) \\
GISSEL ($Z_{\odot}$, no $A_V$ )
             &   &  0.13   &  0.07(81)    &  0.50   &  0.13(19) \\
GISSEL ($Z_{\odot}$, low Lyman blanketing)
             &   &  0.09   &  0.06(81)    &  0.25   &  0.09(27) \\
CWW          &(b)&  0.21   &  0.08(81)    &  0.26   &  0.08(27) \\
CWW $+$ vB   &(c)&  0.17   &  0.08(82)    &  0.26   &  0.07(27) \\
GISSEL ($0.005Z_{\odot}$,$Z_{\odot}$,$5Z_{\odot}$)
             &(d)&  0.10   &  0.05(80)    &  0.30   &  0.11(26) \\
GISSEL ($0.005Z_{\odot}$,$5Z_{\odot}$)  
             &(e)&  0.10   &  0.06(79)    &  0.30   &  0.10(26) \\
GISSEL self-consistent models 
             &(f)&  0.09   &  0.05(80)    &  0.26   &  0.10(27) \\
\hline 
\end{tabular}
\caption{Summary of dispersions in the HDF $z_{\rm phot}$ measurements
for the different sets of templates, where $\delta_z = \sigma/\left[
1+\left< z \right> \right]$, using the Calzetti's law with $A_V$
ranging from 0 to 1.2 magnitudes.  The total number of objects
considered in each non-catastrophic sample is given in brackets. See
more details in text.  }
\label{tabdelta}
\end{center}
\end{table*}

In general, the reasons of failures can be ascribed to many effects,
such as a wrong photometry (systematic errors when measuring
magnitudes or underestimated photometric errors) leading to a highly
unlikely fit, or a probability function with significant secondary
peaks, because of degeneracy among the fit parameters, or a relatively
``flat'' probability function due to a lack of sufficient photometric
information.  
The last explanation applies particularly to the object at $z_{\rm
spec}=5.64$, which is detected only in filter F814W and which is at
the limit of detection in F450W, with $S/N\simeq 1.5$.
However, if we use all the available photometry, disregarding the 
$S/N$ criterion, we obtain $z_{\rm phot}=5.13$. 
The object at $z_{\rm spec}=2.931$ is placed at low redshift by other
groups (Fern\'andez-Soto et al. 1999, Arnouts et al. 1999).
Nevertheless a secondary peak, with a very small $\chi^2$ probability,
is found at $z=2.90$.

We can remark that at high redshift the cases (b) and (c) are better
centred around the spectroscopic value. However, their $\chi^2$
values are higher than in case (a). The reason suspected for that is
the one-to-one relation introduced here between the Lyman-forest
absorption and the redshift.  
We investigate this problem by assigning different values to the
Lyman-forest decrement, multiplying the values of the mean line
blanketing $\left< D_A \right>$ and $\left< D_B \right>$ provided by
Madau (1995) by a factor 0.5 and 1.5, then increasing or decreasing
the absorption (Furusawa et al. 2000).  We found a better fit to the
HDF data when the Lyman forest along the line of sight produces a
smaller flux decrement with respect to the mean value.
In this case we obtain $\delta_z = 0.09$ for the GISSEL case (a),
a value which is similar to the value of CWW SEDs.
An overestimate of absorption due to neutral hydrogen induces a
subsequent and systematic underestimate of redshifts, because the same
attenuation of the flux could be reproduced with a solution at lower
redshift.
Hence a careful knowledge of the UV region of SEDs is essential to
accurately assess $z_{\rm phot}$; furthermore, the Lyman forest
represents the most important signature of spectra in the high
redshift regime.  Thus it is important to allow the blanketing in the
Lyman forest to span a sufficiently wide range of values in order to
prevent systematic effects at high-$z$, which could depend on the line
of sight.

It is worth to notice that, even if all the template SEDs reproduce
the spectroscopic redshifts on the HDF with sufficient accuracy, the
redshift distributions of galaxies could change significantly when we
are dealing with objects fainter than the spectroscopic limits, for
which no training set is available. 
When the redshift distribution obtained on the HDF with CWW templates
is compared with the equivalent one computed with GISSEL templates,
there are no strong differences in the overall distribution.
Nevertheless, this result could not apply to all cases. 
A straightforward example is the case of a deep photometric survey
using visible filters only, without near-IR photometry, and designed
to probe the low surface-brightness regime.
It is easily shown that, in this case, a degenerate solution could
exist for the faintest ``blue'' sources, for which it is impossible to
decide between a low-$z$ solution (low surface-brightness object with
a very young stellar population, as presented in next subsection) and
a relatively bright $1<z<2.5$ galaxy, with ongoing star-formation (no
strong signatures on a continuum increasing bluewards).
In that case, using the CWW templates alone will tend to select the later
solution systematically, whereas including templates spanning a wide
range of ages for the stellar population (such as GISSEL) could select
the former solution, thus leading to a completely different redshift
distribution.
We prefer to adopt a relatively large number of GISSEL's templates,
to supply a wide baseline for modeling the age effects, rather than
to assume the evolution reproduced by the transformation in a
different local spectral type.

\subsection{Age of the stellar population}

   Photometric redshifts are efficient when a spectral feature is
detected through the filters with an important strength as compared to
photometric uncertainties. When we are dealing with the stellar
continuum of a young stellar population, the $4000$\,\AA\ break
becomes visible at $\sim 10^{7}$ years (see Bruzual \& Charlot
1993). In most cases, this lack of strong features could not be
compensated by the presence of strong emission lines, simply because
such lines have a negligible effect on the integrated energy when
using broad-band filters (see Section~\ref{elines}).

\begin{figure*}
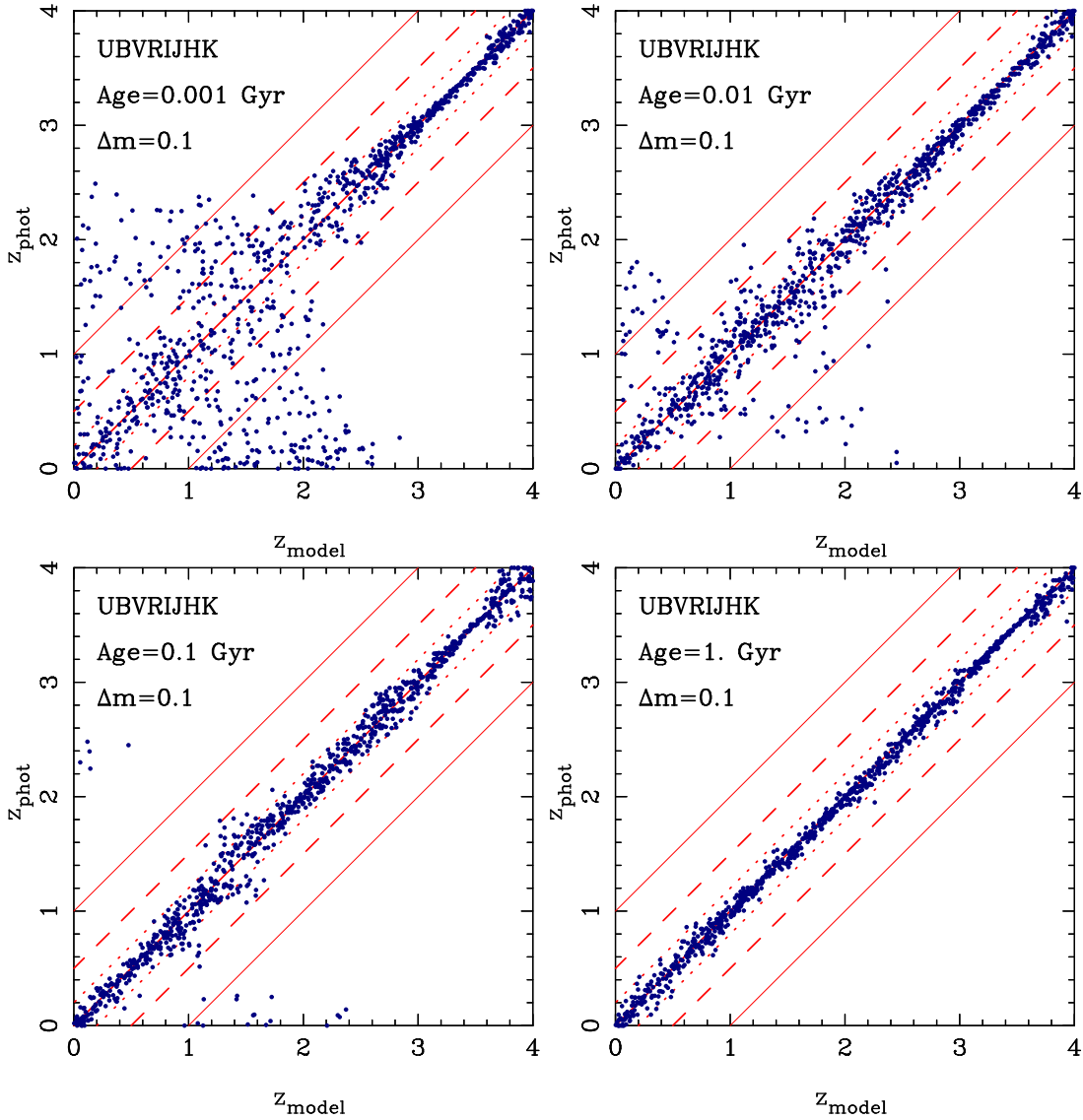

{\centering \leavevmode
\psfig{file=9769_f6a.ps,width=.40\textwidth}
\psfig{file=9769_f6b.ps,width=.40\textwidth}
\newline
\psfig{file=9769_f6c.ps,width=.40\textwidth}
\psfig{file=9769_f6d.ps,width=.40\textwidth}
}
\caption{Comparison between $z_{\rm model}$ and $z_{\rm phot}$ for simulated 
catalogues of single burst galaxies with $\Delta m =0.1$, 
filters set {\em UBVRIJHK\/} and ages of galaxies
$10^6, 10^7, 10^8$ and $10^9$\,yr. }
\label{fig_age}
\end{figure*}

In order to study the effects of age on $z_{\rm phot}$ estimates as a
function of redshift, we have produced different sets of catalogues
corresponding to different ages, all of them with a uniform
distribution in $z$ for the delta burst SED (single stellar population
model).
Figure~\ref{fig_age} displays the general trends of $z_{\rm phot}$
versus $z_{\rm model}$ for representative ages and the {\em
UBVRIJHK\/} set of filters.
In this case the set of templates used is the basic GISSEL one
with solar metallicity. 
At $z_{\rm model}\ga 3$, the redshift determination is accurate for
any age because of the presence of Lyman break in the filter $U$.  At
smaller redshifts, $z_{\rm phot}$ is based on the 4000\,\AA\ break as the
strongest spectral signature, and it is visible only in systems which
are a few $\sim 10^{7}$ years old. 

\begin{figure}[t]
{\centering\psfig{file=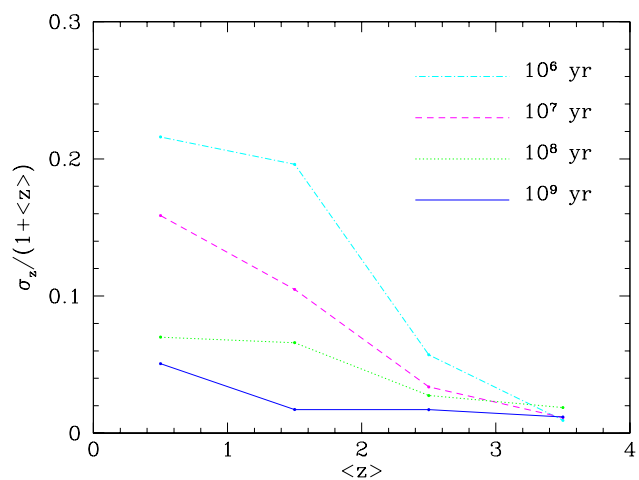,width=0.49\textwidth,angle=270}}
\caption{The dispersion $\sigma_z/(1+\left< z \right>)$ as a function 
of the mean redshift of the considered range. Ages are $10^6, 10^7, 
10^8, 10^9$ and $10^{10}$\,yr from top to bottom.}
\label{sigmaz_age}
\end{figure}

The results obtained applying {\it hyperz\/} to these catalogues are
summarized in Figure~\ref{sigmaz_age}, where we show the effect
described above by means of the dispersion in four redshift bins: the
value of $\sigma_z/(1+\left< z \right>)$ decreases increasing the
redshift and the age of the stellar population. 

\subsection{Cosmology}

   The effects of cosmological parameters ($H_0$, $\Omega_0$ and
$\Omega_{\Lambda}$) are only related to the age allowed to the stellar
population at a given redshift. When using {\it hyperz\/}, the age of
the stellar population can be optionally limited to the age range
permitted by the cosmological parameters. In order to quantify such
effect on $z_{\rm phot}$, if any, we have compared the results
previously obtained on the HDF (with the crude age limitation given
above) with those obtained without age constraints, and also with a
different set of cosmological parameters ($\Omega_0=0.3$,
$\Omega_\Lambda=0.7$ and $H_0=50\,{\rm km \,s^{-1}\,Mpc^{-1}}$).
These results show that the effect of the cosmological parameters on
the $z_{\rm phot}$ estimate is negligible, because they affect
$\delta_z$ by less than $1$\%.

\subsection{Reddening}
\label{red}

\begin{figure}[t]
\centering\psfig{file=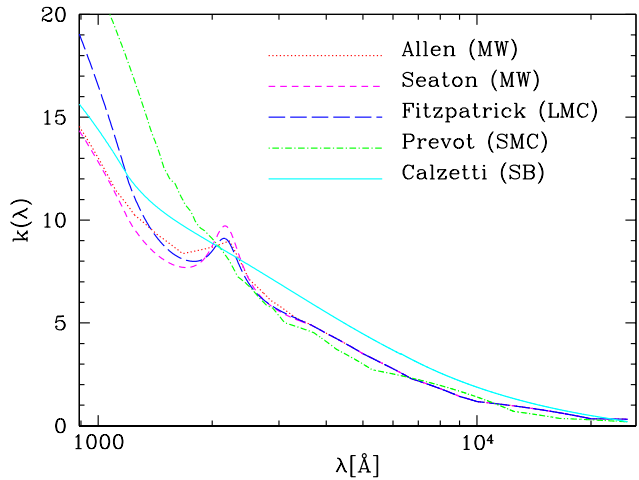,width=0.49\textwidth,angle=270}
\caption{Extinction curves $k(\lambda)$ for the different 
reddening laws implemented in {\it hyperz\/}.}
\label{figred}
\end{figure}

The five reddening laws presently implemented in {\it hyperz\/} are: 
\begin{enumerate}
\item Allen (1976) for the Milky Way (MW);
\item Seaton (1979) fit by Fitzpatrick (1986) for the MW;
\item Fitzpatrick (1986) for Large Magellanic Cloud (LMC);
\item Pr\'evot et al. (1984) and Bouchet et al. (1985) 
for Small Magellanic Cloud (SMC);
\item Calzetti et al. (2000) for starburst galaxies. 
\end{enumerate}
The different laws are presented in Figure~\ref{figred}.

Recent studies on high redshift galaxies and star formation obscured
by dust have shown the importance of reddening in the high-$z$
universe.
In order to probe this issue on $z_{\rm phot}$ computations, we have
compared the results previously obtained on the HDF to those obtained
assuming no reddening, all the other parameters being fixed.
We found $\delta_z = 0.13 (0.07$ without catastrophic objects) for the
low-$z$ bin and $\delta_z = 0.50 (0.13)$ for the high-$z$ one, but
with a much higher percentage of catastrophic identifications: 10
objects at $z_{\rm spec}\simeq 3$ are erroneously identified as low
redshift galaxies.

Therefore, keeping a wide range of reddening values seems to be
essential to reproduce the SEDs of high redshift galaxies. 
According to Steidel et al. (1999), the typical $E_{B-V}$ for galaxies
to $z \sim 4$ is $0.15$ mags, thus $A_V \simeq 0.6$ mags when using a
Calzetti's law. The maximum $A_V$ allowed in our calculations is about
2 times this value.
 
Moreover, we conducted a test to study the influence of the different
reddening laws, using all the implemented possibilities.  
We found that the laws reproducing the extinction of the Milky Way and
the Large Magellanic Cloud are not appropriate to fit the SEDs of high
redshift galaxies ($z_{\rm spec}>2$), whereas they leave the low
redshift region unaffected.
Instead, the fourth law, corresponding to the Small Magellanic Cloud,
produces results similar to those obtained with the curve provided by
Calzetti et al. (2000). It correctly assigns the $z_{\rm phot}$ to the
high redshift objects, but it places a couple of low $z_{\rm spec}$
objects at higher $z_{\rm phot}$. The last effect is probably due to
the higher and steeper $k(\lambda)$ at short wavelength as compared to
Calzetti's, which mimics the additional effect of the UV attenuation
induced by the Lyman forest.
At high redshift, the most important wavelength region is the UV,
between $1000$\,\AA\ and $3000$\,\AA , where the considered laws give
quite different trends, thus modifying in a different way the
magnitudes and producing different values of $\chi^2$.
In fact, most of the fits to the HDF sample using reddening laws from 1 to
4 produce worse $\chi^2$ values than the Calzetti's law, in particular
for those objects requiring $A_V >0.6$.  These galaxies cannot be
reproduced by the MW and LMC laws, even when the limit of $A_V$ is
increased up to $A_V=2$.

Thus, the slope of the selected reddening law at short wavelengths
must be defined carefully; the extrapolation used here to extend the
laws 1 to 4 towards wavelengths not covered by data is rather poor.
These considerations get stronger evidence that the modeling of the UV
region of SEDs is essential to recover correctly the high $z$
galaxies.
The re-emission of energy coming from dust heated by massive star
formation does not affect the present results, because we concentrate
on the UV to near-IR bands.

\subsection{Metallicity}

   We have also checked the influence of the metallicity on the 
$z_{\rm phot}$ estimates using the same HDF training sample. The same
computations have been done using different and extreme assumptions
for the metallicity of the stellar population, with values ranging
from $0.005 Z_{\odot}$ to $5 Z_{\odot}$ (as allowed by Bruzual \&
Charlot's models). We have also developed a self-consistent set of
templates, where the evolution in metallicity of the stellar
population is explicitly taken into account (cf. Mobasher \& Mazzei 1999).
In other words, there is a natural link between the age of the stellar
population and its mean metallicity. For all metallicity cases, we
have built up the same closed-box systems presented before: a constant
star-forming galaxy and six $\mu$-models.

Three sets of templates were considered: the 3 different metallicities
together (solar and the 2 extreme values), the two extreme values
alone, and the self-consistent model.  
A comparison among all these cases is given in Figure~\ref{hdf}
(d,e,f).  The dispersions at low redshift without failed objects are
$\delta_z = 0.05,0.06,0.05$ respectively, for the 3 different sets. At
high redshift we found $\delta_z = 0.11,0.10,0.10$, under the same
assumptions.
A slight improvement on the accuracy of $z_{\rm phot}$ at $z \la 1.5$
is observed when several different metallicities are used together,
and the self-consistent model (f) produces the best fit in this
redshift range.  On the other side, including different metallicities
does not affect the high redshift determinations.

\subsection{Initial Mass Function}
\label{imf}

The influence of the IMF has also been tested on the HDF spectroscopic 
sample.  We have used the self-consistent modeling, which takes into
account the evolution in metallicity of the stellar population and
produces the best fit to the HDF data when using the Miller \& Scalo
IMF (1979).
We have built up the same closed-box models for 2 additional IMFs,
Salpeter (1955) and Scalo (1986), keeping the same upper mass limit
for star formation. When applying these new templates to the HDF
sample, we find exactly the same results in terms of $z_{\rm phot}$
accuracy.
Looking more carefully to the results obtained for individual objects,
we find that the $z_{\rm phot}$ estimates are approximatively the same,
whatever the IMF used. This result is easy to understand because the
changes induced on the stellar continuum by the different IMF slopes
are compensated in most cases by the other parameters (reddening, age,
...), thus giving the same $z_{\rm phot}$ result but a different solution
in the parameter space. 

When we compute $z_{\rm phot}$ on simulated data, the $z_{\rm phot}$
accuracy is the same when we use a unique IMF in model galaxies and
templates and when we use a different IMF in both settings. 
In addition, we have checked on possible systematic changes
on the spectral types derived by {\it hyperz\/} in the later case,
with negative results. In particular, a model catalogue built with
Miller \& Scalo IMF was analysed with Salpeter and Scalo IMFs, and the 
results were the same as in the Section~\ref{param_rec} below.
This strengthens the idea of the IMF being a secondary parameter in
$z_{\rm phot}$ estimates.

\subsection{Emission lines}
\label{elines}

    As long as we are dealing here with broad-band photometry, the
presence of emission lines on the spectra has a relatively small
effect on the integrated fluxes, and thus a small influence on the
$z_{\rm phot}$ results. This can be easily quantified when we consider the
sample of blue compact galaxies at $z \le 1.4$ studied by Guzm\'an et
al. (1997), and the samples of star-forming galaxies described by
Cowie et al. (1995), Glazebrook et al. (1995) and Terlevich et
al. (1991).  
At relatively low redshift, the main emission lines to consider are
[O{\sc ii}]$\lambda 3727$, H${\alpha}$, H${\beta}$ and
[O{\sc iii}]$\lambda\lambda 4959,5007$, [O{\sc ii}] and H${\alpha}$ 
being the most important contributions to the integrated fluxes.
According to Guzm\'an et al. (1997), the [O{\sc ii}]$\lambda 3727$
luminosity of star-forming galaxies can be approximated by 
$L(\mbox{[O{\sc ii}]}) \sim 10^{29} W_{\mbox{[O{\sc ii}]}} L_B$, where 
$W_{\mbox{[O{\sc ii}]}}$ is the equivalent width and $L_B$ is the blue 
luminosity in solar units.

For our purposes, an emission line can be overlooked when
$f(\mbox{e-line})/f_{\lambda} \le 1 - 10^{-0.4 \Delta m}$, where
$f(\mbox{e-line})$ and $f_{\lambda}$ are, respectively, the integrated
fluxes within the emission line and the stellar continuum through the
filter, and $\Delta m$ is the photometric uncertainty in magnitudes.
A realistic value of $\Delta m \sim 0.05$ to 0.1 mags ($\sim 5 $ to
10\% uncertainty) imposes $f(\mbox{e-line})/f_{\lambda} \le 0.05$ to
0.1. The limit in equivalent width for galaxies in the Guzm\'an et
al. sample is a few times $100$\,\AA, thus most compact star-forming
galaxies fulfill this condition. Even when we consider the typical
luminosities of vigorous star-forming sources ($L({\rm [O\mbox{\sc ii}]}) 
\sim 10^{42}$ erg/s, Cowie et al. 95, Glazebrook et al. 1995), emission 
lines  are found to be negligible in most of them.
Also the large majority of H{\sc ii} galaxies in the Terlevich et al.
(1991) local sample fulfill the condition.

Thus, emission lines do not seem to influence significantly the
$z_{\rm phot}$ results on star-forming galaxies.
On the contrary, this is not the general case when we are dealing with
AGNs, or when the photometry is obtained through narrow-band filters.  
We have not considered here neither the contribution of AGN to
the simulated samples, nor the influence of such templates on the
final accuracy when we are dealing with real data. AGN SEDs could be
easily introduced in our present scheme, and this particular
application is presently under development (Hatziminaoglou et
al. 2000).

\subsection{Recovering the main SED parameters through {\it hyperz \/}}
\label{param_rec}

\begin{figure}[t]
\centering\psfig{file=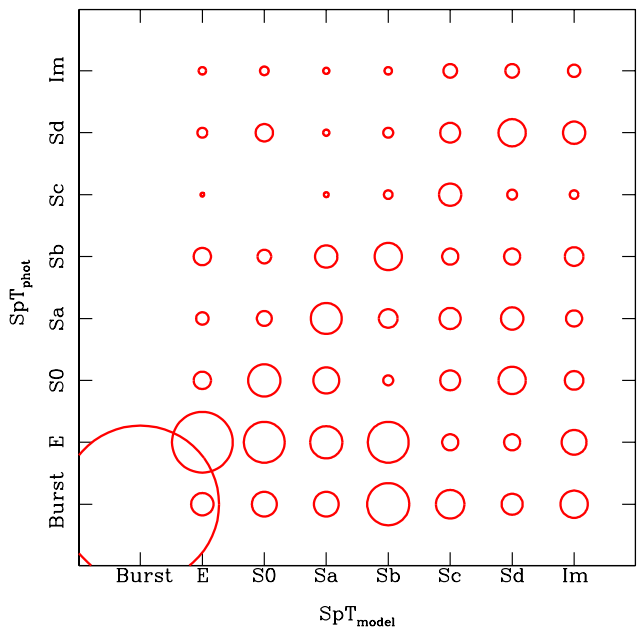,width=0.49\textwidth}
\caption{Comparison between the model spectral types and the best-fit
templates recovered by {\it hyperz\/}, for the simulations computed in
Section~\ref{simul}.  Spectral types, from 1 to 8, correspond to
galaxies ranging from early (Bursts/E) to late (Im) types. Circle
sizes scale with the number of objects. Ideally only the diagonal
region should have been populated.}
\label{fig_stypes}
\end{figure}

  As mentioned before, {\it hyperz \/} allows to obtain the $z_{\rm
phot}$ and the best fit parameters across the whole space.  The
fitting procedure does not favour any parameter in particular.  The
homogeneous simulations presented in Section~\ref{simul} could be used
to briefly discuss on the efficiency to recover the most relevant
input parameters: the spectral types, the age of the stellar
population and $A_V$.  
Because of the degeneracy between these parameters, and the lack of
spectral resolution, we only expect a rough spectral type to be
retrieved from broad-band photometry.  We have considered the 8
spectral types presented in Section~\ref{method} to illustrate the
case. A general trend appears when comparing the model and retrieved
spectral types, whatever the redshift, filter combination and
photometric accuracy, with single bursts and early types being more
easily identified than late types at all redshifts.
Figure~\ref{fig_stypes} displays an example obtained with the {\em
UBVRIJK} filter combination and 10\% photometric accuracy, excluding
catastrophic identifications (less than 1\% in this case).  The trend
remains the same whatever the distribution in types, from these
detailed 8 types to a rough Burst-E/S/Im distribution.  Lowering the
$S/N$ or the number of filters slightly increases the trend in terms of
contrast between the early type and late type behaviour. Late type
misidentifications are due to the degeneracy between age of the
stellar population and spectral type, such galaxies being incorrectly
assigned to younger {\it and} earlier types.  In other words, there is
often a burst-like template, of suitable age and length, which is able
to fit the dominant stellar population of a galaxy observed through
broad-band filters.  The results are the same whatever the
configuration in the parameter space, in particular, changing the
order or the position of the different templates in the space produces
the same results.  Degenerate solutions in the redshift dimension are
systematically displayed by {\it hyperz}, but this is only an
option for the other dimensions of the parameter space.  There is no
systematic trend in the case of catastrophic identifications, but more
than 90\% of such objects in these simulations have misidentified
spectral types as well.

In the case of $A_V$, the procedure will choose the best and the
lowest possible value. The results in this case are much better,
whatever the $S/N$, provided that near IR filters are included. Using
a grid of $\Delta A_V = 0.2$ to explore the parameter space, the
typical value of $\sigma_{AV} = \sqrt{\sum
(\Delta_{AV}-\left<\Delta_{AV} \right>)^2/(N-1)}$ ranges between
$\sigma_{AV} = 0.15$ and $0.3$, for photometric accuracies between 5 and
30\%, for all the filter combinations including $J$, $H$ or $K$ (or a
combination of them).  In all the other cases, $\sigma_{AV} = 0.3$ to
0.45, for photometric accuracies between 5 and 30\%.  These values are
an average through all the spectral types and redshifts, excluding
catastrophic identifications. Similar estimates on catastrophic
objects show an increase between $+0.05$ and $+0.3$ on
$\sigma_{AV}$, depending on the filter set.

In summary, it is difficult to obtain detailed information on the
spectral types from broad-band photometry alone, and this is probably
the result of the poor spectral resolution.  Near IR photometry allows
to constraint the $A_V$ value for all spectral types.  Only early type
galaxies could be reliably identified by this method.  For later
types, only a rough estimate of the SED type could be obtained, in
terms of ``blue'' or ``red'' continuum.  The classification in this case
shall either include the spectral type {\it and} the age of the
stellar population, or be based on a simple set of templates such as
CWW.  

\section{Expected accuracy on real data}
\label{realsim}

    In order to discuss on the expected accuracy and possible
systematic errors when exploring real data, we have performed a
complete set of simulated catalogues, with a more realistic (non
uniform) redshift distribution and $S/N$ ratio along the SEDs.  We
have adopted the simple PLE model proposed by Pozzetti et al. (1996,
1998), with minimal changes, to derive the redshift distributions and
to assign a magnitude to each object in the different filters.  Four
galaxy types  (exponentially decaying SFR with characteristic time
$\tau = 1$\,Gyr and $10$\,Gyr, constant SFR with evolution in time and
at a fixed age of $0.1$\,Gyr) and their corresponding luminosity
functions are used to reproduce the number of galaxies expected at a
given redshift and absolute magnitude $M_{b_{\rm J}}$.
Apparent magnitudes are computed from the evolved SEDs, with ages
depending on the redshift considered, the formation redshift, set to
$z_{\rm form}=7$, and the cosmological parameters.  Photometric errors
are scaled to apparent magnitudes assuming the approximate relation
$\Delta m \simeq 2.5 \log [1+1/(S/N)]$, where $S/N$ is the signal to
noise ratio, which is given as a function of the apparent magnitude
through $S/N = (S/N)_0 10^{-0.4(m-m_0)}$, $(S/N)_0$ being the signal
to noise ratio at a given reference magnitude $m_0$. For simplicity,
the photometric error is assigned to the apparent magnitude $m$
according to a Gaussian distribution of $\Delta m$ fixed $\sigma$.
This relation is set to reproduce the rapid increase of uncertainties
when approaching the limiting magnitudes.  According to these
equations, a value of $S/N=10$, corresponding to $\Delta m \simeq
0.1$, is reached $2.5$ magnitudes brighter than the magnitude
corresponding to $S/N=1$.  An object with $S/N<1$ is non-detected in
the involved filter ($m=99$).  An object is included in the final
catalogue if it is detected in the filter $I$ (assuming that this is
the selection filter), and in at least two other filters. The last
requirement is needed to compute $z_{\rm phot}$.

\begin{table}[t]
\begin{center}
\begin{tabular}{ccc|cc}
\hline
Filter  & $m_{\rm lim}$(d) & $m_{\rm lim}$(s) & Filter & $m_{\rm lim}$\\
\hline \hline
$U$     & 29.0             & 25.5             & F300W  & 29.0 \\
$B$     & 30.0             & 26.5             & F450W  & 30.5 \\
$V$     & 29.5             & 26.0             & F606W  & 30.0 \\
$R$     & 29.5             & 26.0             & F814W  & 29.0 \\
$I$     & 28.5             & 25.0             & $J$    & 25.0 \\
$Z$     & 27.5             & 24.0             & $H$    & 24.0 \\
$J$     & 25.0             & 21.5             & $K'$   & 23.5 \\
$H$     & 24.0             & 20.5             &        & \\
$K$     & 23.5             & 20.0             &        & \\
\hline 
\end{tabular}
\caption{Limiting magnitudes at $S/N=1$: (d) deep pencil beam-like survey, 
(s) shallow ground-based survey.}
\label{tabmlim}
\end{center}
\end{table}

The same filter combinations discussed in Section~\ref{simul}
have been used to produce the new simulated catalogues. The
simulations in Section~\ref{simul} represent an ideal case, with an
infinite depth and a fixed photometric error, disregarding the
dependence on errors versus magnitudes. However, the relevant
quantities $\sigma_z$, $l$\% and $g$\% strongly depend on the number
of objects in each redshift bin and then on the limiting magnitudes.

To give a qualitative idea of the accuracy expected with different
observational configurations, we consider two representative cases. 

\subsection{Deep pencil beam surveys}

Firstly, we focus on simulations obtained in the case of a pencil
beam-like survey, i.e. a very deep observation, covering a small area.  
From the photometric point of view, the main improvement with respect
to the uniform distributions presented above is that we can introduce,
for each object, a realistic $S/N$ in the different filters, with
different values from filter to filter.
We assume that the detection limit is reached ($S/N=1$) at magnitudes
similar to the limiting magnitudes of the HDF, as reported in the 
column $m_{\rm lim}$(d) of Table~\ref{tabmlim} and in the right part 
of the same table for the HDF-N filters. 
To obtain approximately the same number of galaxies observed in the
HDF, a field of 5\,arcmin$^2$ has been simulated. 
The percentages of spectral types included in the simulated
catalogue are $\sim 11, 58, 28, 3$\% for E, Sb, Im, and Im(${\rm age}
= 0.1$\,Gyr) respectively.
In order to reproduce the observed number counts at faint magnitudes
(Williams et al. 1996) we assume an open cosmological model, with
$\Omega_0=0.1$ and $\Omega_\Lambda = 0$.
In this case, the peak of the redshift distribution is at $z\ga 1$
and very few objects are seen at low-$z$, 
in particular at redshifts between $z=0$ and $z=0.4$. 
Moreover, the PLE model is known to overestimate the population of high 
redshift galaxies. 

In Table~\ref{newgain} we display the computed quantities
$\sigma_z$, $l$\% and $g$\% for the set of filters of the HDF-N, and
for all the other deep survey combinations considered in
Section~\ref{simul} (marked by (d) in the second column).  The table
contains the dispersion and the percentages of spurious and
catastrophic objects, computed from a set of 10 independent
simulations for each configuration.
The interpretation of data in Table~\ref{newgain} must take into
account that the definition of $g$\% depends on the dispersion
$\sigma_z$ computed using the correctly assigned objects, and this
quantity is quite sensitive to the different filter sets and redshift
bins.
Nevertheless, these simulations take properly into account the
observed properties of galaxies in deep surveys, such as the presence
of faint objects with huge photometric errors, and the lack of
detection in some filters leading to an uncertain $z_{\rm phot}$
estimate (that is, increasing the probability of misidentifications,
enlarging the error bars and the dispersion around the true value).
In particular, this effect is evident when looking at the trend
of the $l$\% values. At higher redshift we find an increasing number
of faint objects that are non detected in some filters. This lead to
an increase of $l$\%. Because of the depth of limiting magnitudes, we
adopt the non detection law number 1 for optical filters and 2 for the
near infrared ones.
In the highest redshift bins, the increase of the dispersion value
tends to mitigate the effect of the $z_{\rm phot}$ deterioration in
the value of $g$\%.

\begin{figure}[t]
\psfig{file=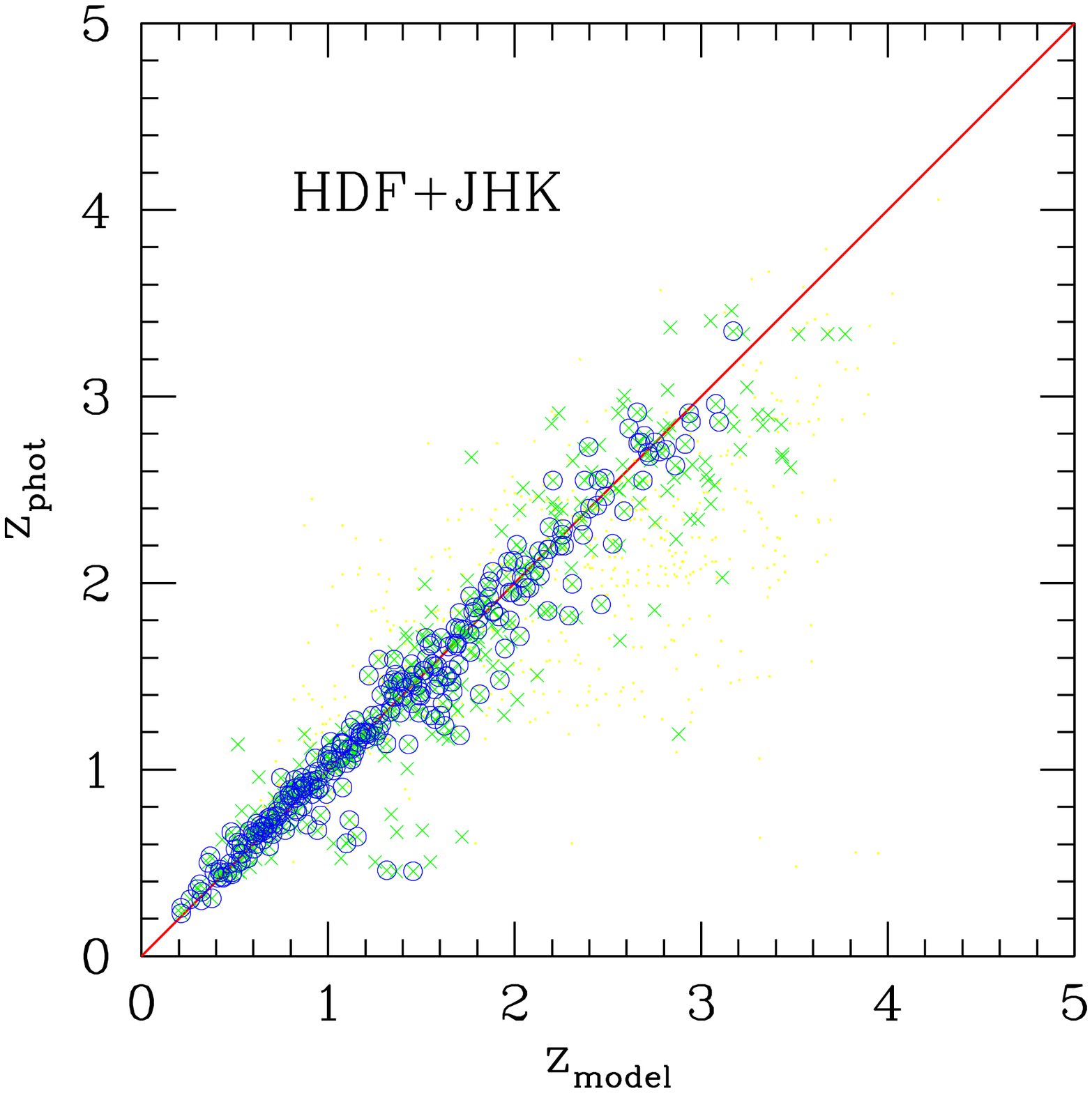,width=.49\textwidth}
\psfig{file=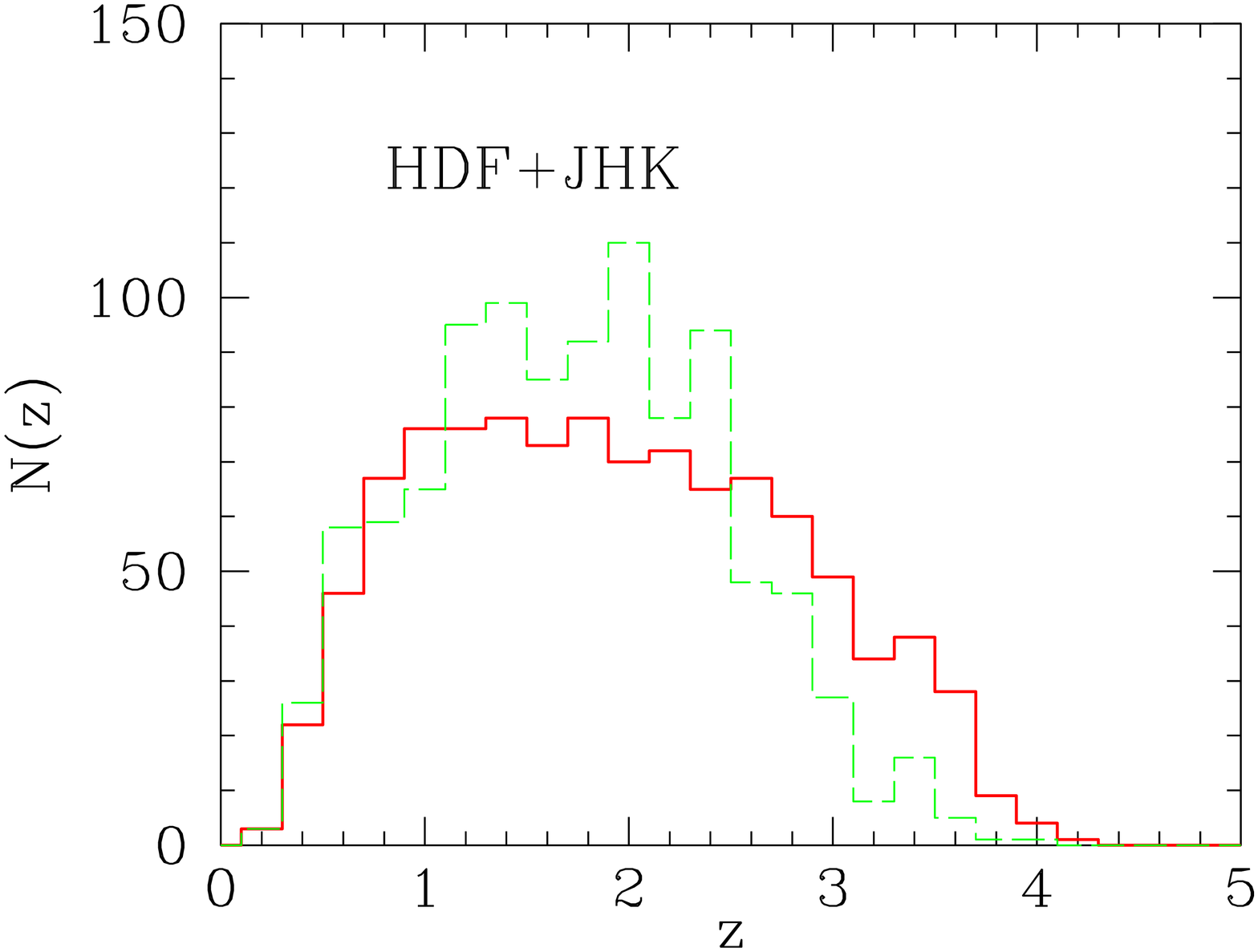,angle=270,width=.49\textwidth}
\caption{Top: comparison between $z_{\rm model}$ and $z_{\rm phot}$
for realistic catalogue HDF-like. See the text for the considered
limiting magnitudes. Small dots, crosses and circles correspond to
objects brighter than $(S/N)_{\rm lim}=1, 5, 10$ respectively, at
least in the $I$ filter and in two other filters. Bottom: redshift
distributions for the simulation on the top with $(S/N)_{\rm
lim}=1$. Solid line: $N(z_{\rm model})$. Dashed line: $N(z_{\rm
phot})$.}
\label{figreal1}
\end{figure}

In the case of HDF-N filter set, we considered also two
subcatalogues built with more restrictive selection criteria,
requiring the detection both in $I_{814}$ {\it and} in at least two
other filters to be $S/N \ge 5$ and $S/N \ge 10$.  Statistics
concerning these simulations are tabulated in Table~\ref{newgain}.
Obviously, when considering objects with increasing $S/N$, the
accuracy of $z_{\rm phot}$ estimate significantly improves.  In
Figure~\ref{figreal1} we show the results obtained on the comparison
between $z_{\rm phot}$ and model redshifts, and also on the $N(z_{\rm
model})$ versus $N(z_{\rm phot})$ reconstruction. Most of the
discrepancy is due to objects with $S/N \le 10$.

\subsection{Shallow wide field surveys}

\begin{figure}[t]
\psfig{file=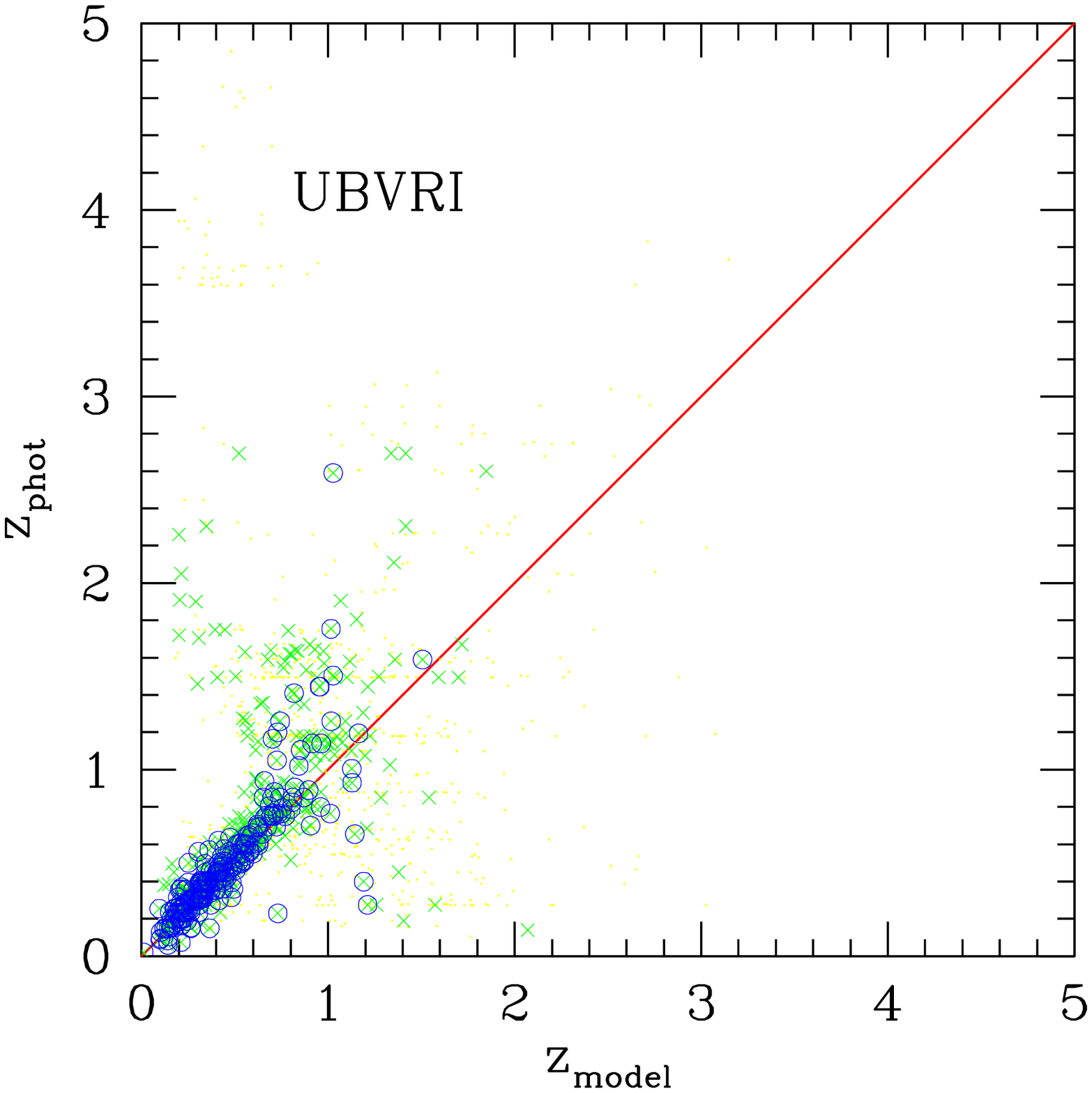,width=.49\textwidth}
\psfig{file=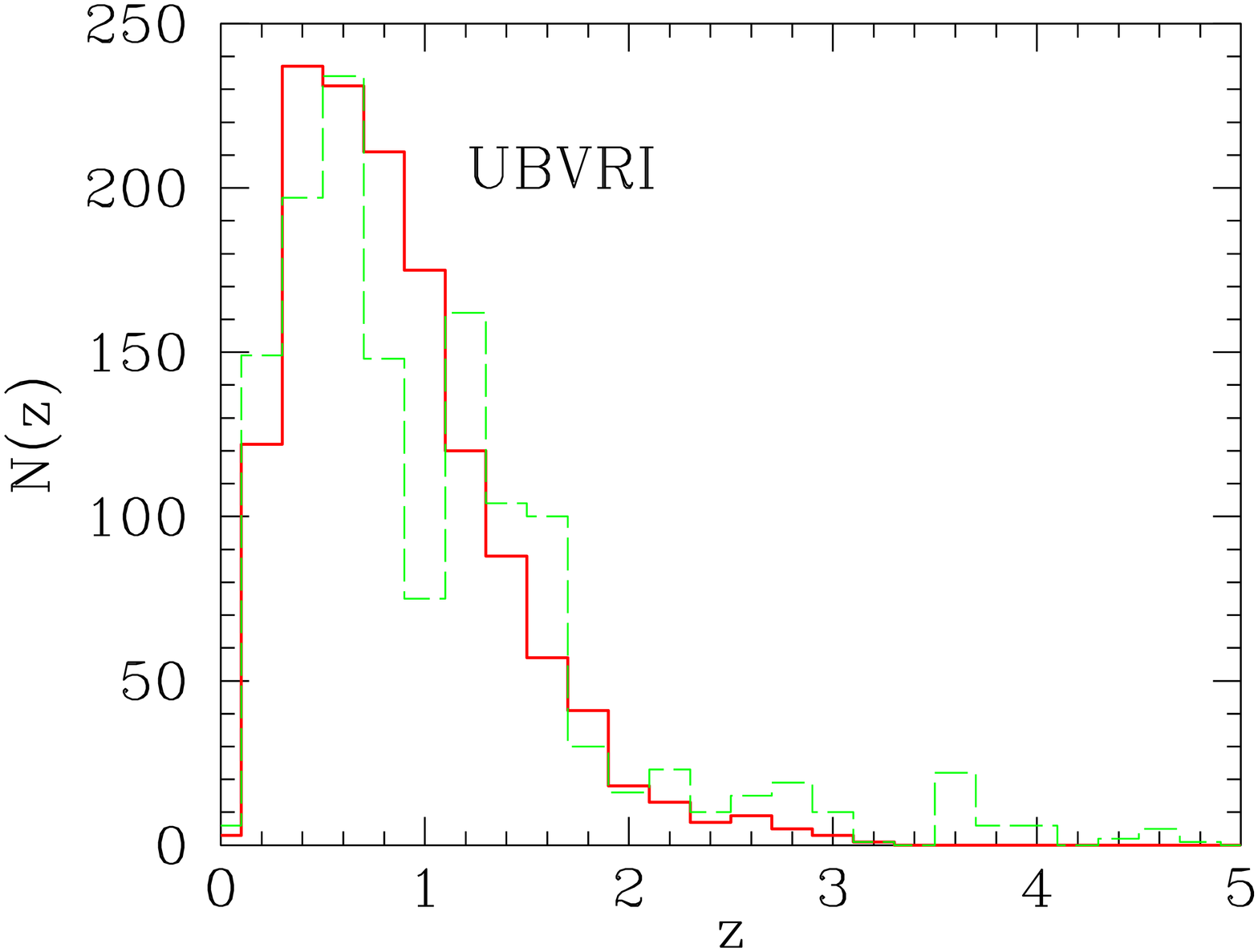,angle=270,width=.49\textwidth}
\caption{Top: comparison between $z_{\rm model}$ and $z_{\rm phot}$
for realistic catalogue for a shallow survey with 5 filters. The
symbols are the same as in Figure \ref{figreal1}. Bottom: redshift
distributions for the simulation on the top with $(S/N)_{\rm
lim}=1$. Solid line: $N(z_{\rm model})$. Dashed line: $N(z_{\rm
phot})$.}
\label{figreal2}
\end{figure}

\begin{figure}[t]
\psfig{file=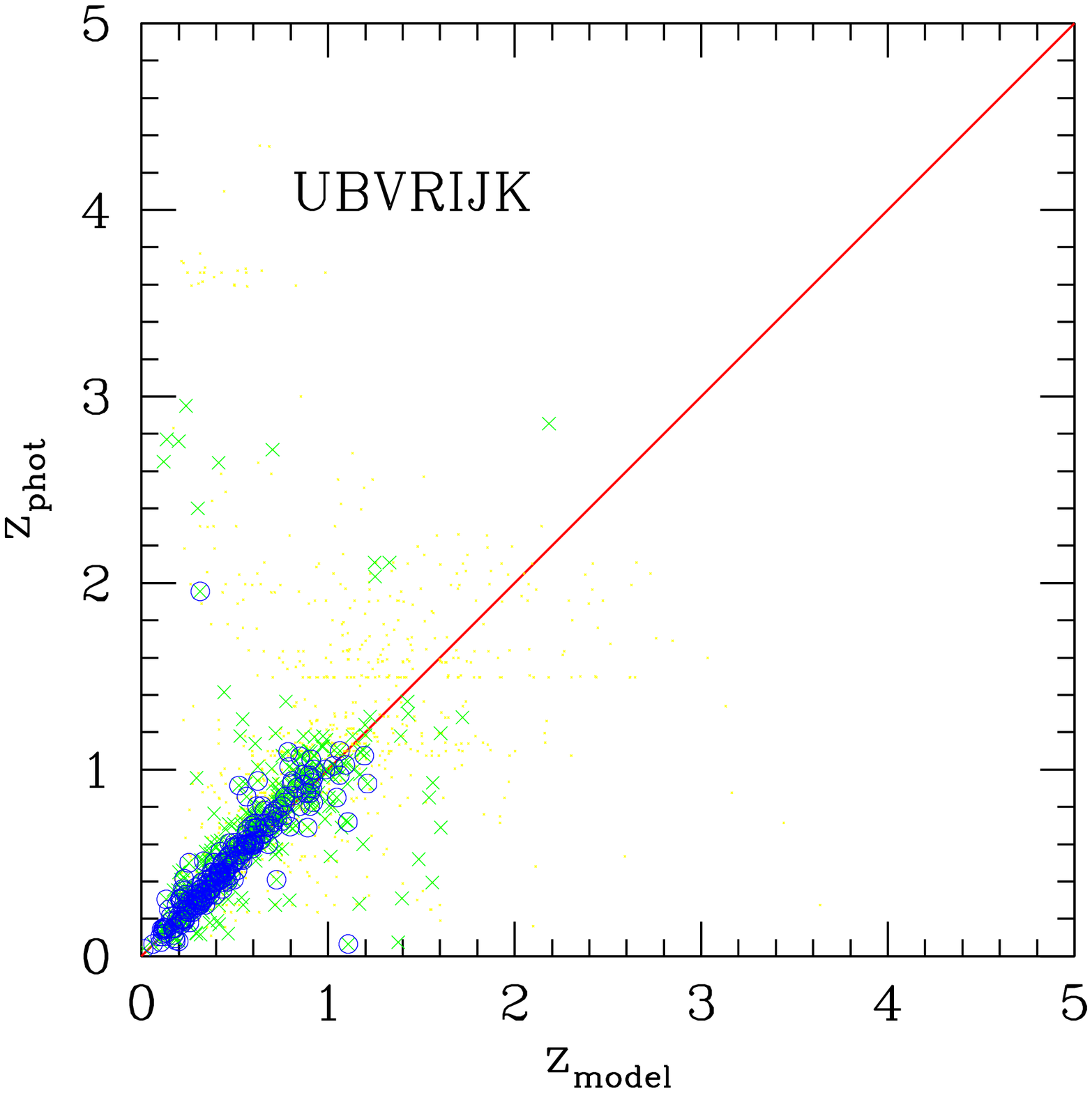,width=.49\textwidth}
\psfig{file=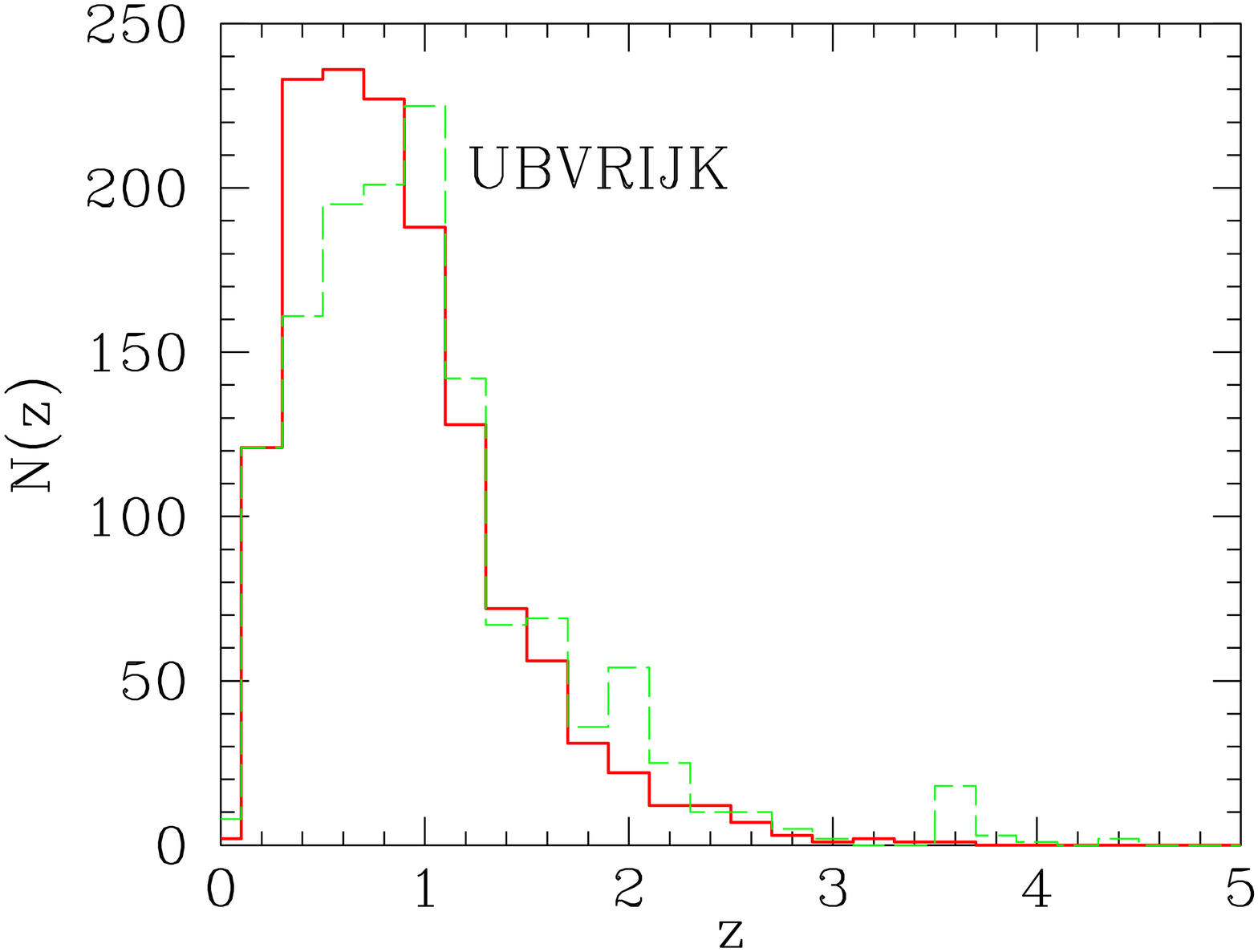,angle=270,width=.49\textwidth}
\caption{Top: comparison between $z_{\rm model}$ and $z_{\rm phot}$
for realistic catalogue for a shallow survey with 7 filters. The
symbols are the same as in Figure \ref{figreal1}. Bottom: redshift
distributions for the simulation on the top with $(S/N)_{\rm
lim}=1$. Solid line: $N(z_{\rm model})$. Dashed line: $N(z_{\rm
phot})$.}
\label{figreal3}
\end{figure}

\tabcolsep 0.07cm
\begin{table*}[t]
\begin{center}
\begin{tabular}{|c|c|ccc|ccc|ccc|ccc|ccc|}
\hline
 & &
\multicolumn{3}{c|}{$z = 0.0$ -- 0.4} & \multicolumn{3}{c|}{0.4 -- 1.0} &
\multicolumn{3}{c|}{1.0 -- 2.0} & \multicolumn{3}{c|}{2.0 -- 3.0} &
\multicolumn{3}{c|}{$> 3$ }  \\
\cline{3-17}
filters      &   & $\sigma_z$ & $l$\% & $g$\% 
                 & $\sigma_z$ & $l$\% & $g$\%  
                 & $\sigma_z$ & $l$\% & $g$\% 
                 & $\sigma_z$ & $l$\% & $g$\%  
                 & $\sigma_z$ & $l$\% & $g$\% \\
\hline \hline
HDF+{\em JHK}&   & 0.06 & $ 0\pm 0$  & $ 3\pm 7$ 
                 & 0.17 & $ 1\pm 1$  & $ 8\pm 3$ 
                 & 0.29 & $ 1\pm 1$  & $12\pm 2$ 
                 & 0.39 & $ 8\pm 1$  & $ 4\pm 1$ 
                 & 0.34 & $30\pm 3$  & $ 1\pm 1$ \\
$S/N=5$ 
             &   & 0.06 & $ 0\pm 0$  & $ 3\pm 7$ 
                 & 0.09 & $ 0\pm 0$  & $ 8\pm 2$ 
                 & 0.20 & $ 0\pm 0$  & $ 4\pm 2$ 
                 & 0.30 & $ 1\pm 1$  & $ 0\pm 1$ 
                 & 0.30 & $ 2\pm 2$  & $ 0\pm 0$ \\
$S/N=10$ 
             &   & 0.06 & $ 0\pm 0$  & $ 3\pm 7$ 
                 & 0.07 & $ 0\pm 0$  & $ 4\pm 2$ 
                 & 0.15 & $ 0\pm 0$  & $ 2\pm 1$ 
                 & 0.20 & $ 0\pm 0$  & $ 0\pm 0$ 
                 & 0.19 & $ 0\pm 0$  & $ 0\pm 0$ \\
\hline
{\em BVRI}   &(d)& 0.08 & $10\pm 8$  & $38\pm 15$ 
                 & 0.32 & $ 8\pm 2$  & $40\pm 4$ 
                 & 0.46 & $19\pm 1$  & $10\pm 2$ 
                 & 0.56 & $34\pm 3$  & $ 4\pm 1$ 
                 & 0.47 & $31\pm 6$  & $ 7\pm 1$ \\
             &(s)& 0.17 & $30\pm 3$  & $10\pm 2$ 
                 & 0.36 & $11\pm 2$  & $ 7\pm 1$ 
                 & 0.45 & $18\pm 1$  & $ 6\pm 1$ 
                 & 0.70 & $58\pm 7$  & $32\pm 6$ 
                 & ---  & ---        & ---  \\
\hline
{\em UBVRI}  &(d)& 0.06 & $ 0\pm 0$  & $37\pm 13$ 
                 & 0.30 & $ 7\pm 1$  & $21\pm 4$ 
                 & 0.43 & $12\pm 2$  & $10\pm 2$ 
                 & 0.50 & $17\pm 2$  & $ 4\pm 1$ 
                 & 0.46 & $28\pm 3$  & $ 5\pm 1$ \\
$S/N=1$      &(s)& 0.15 & $17\pm 1$  & $25\pm 3$ 
                 & 0.34 & $ 7\pm 1$  & $ 4\pm 1$ 
                 & 0.49 & $19\pm 2$  & $ 2\pm 1$ 
                 & 0.58 & $41\pm 10$ & $14\pm 7$ 
                 & ---  & ---        & ---  \\
$S/N=5$
             &(s)& 0.09 & $ 5\pm 1$  & $ 7\pm 2$ 
                 & 0.26 & $ 2\pm 1$  & $ 2\pm 1$ 
                 & 0.49 & $14\pm 6$  & $ 2\pm 1$ 
                 & ---  & ---        & --- 
                 & ---  & ---        & ---  \\
$S/N=10$
             &(s)& 0.07 & $ 1\pm 1$  & $ 2\pm 2$ 
                 & 0.18 & $ 1\pm 1$  & $ 1\pm 1$ 
                 & 0.43 & $ 4\pm 7$  & $ 4\pm 5$ 
                 & ---  & ---        & ---   
                 & ---  & ---        & ---  \\
\hline
{\em UBVRIZ} &(d)& 0.06 & $ 0\pm 0$  & $20\pm 10$ 
                 & 0.23 & $ 6\pm 2$  & $28\pm 3$ 
                 & 0.45 & $12\pm 2$  & $ 2\pm 1$ 
                 & 0.43 & $12\pm 2$  & $11\pm 1$ 
                 & 0.48 & $24\pm 2$  & $ 6\pm 2$  \\
             &(s)& 0.14 & $18\pm 1$  & $25\pm 2$ 
                 & 0.29 & $11\pm 1$  & $ 7\pm 1$ 
                 & 0.53 & $21\pm 2$  & $ 1\pm 1$ 
                 & 0.42 & $41\pm 9$  & $35\pm 8$ 
                 & ---  & ---        & ---    \\
\hline
{\em UBVRIJ} &(d)& 0.05 & $ 0\pm 0$  & $13\pm 9$ 
                 & 0.19 & $ 3\pm 1$  & $25\pm 3$ 
                 & 0.37 & $ 5\pm 1$  & $ 6\pm 1$ 
                 & 0.42 & $15\pm 2$  & $ 6\pm 2$ 
                 & 0.43 & $28\pm 3$  & $ 2\pm 2$ \\
             &(s)& 0.17 & $11\pm 1$  & $19\pm 2$ 
                 & 0.25 & $ 5\pm 1$  & $ 4\pm 1$ 
                 & 0.45 & $11\pm 2$  & $ 2\pm 1$ 
                 & 0.50 & $35\pm 8$  & $28\pm 7$ 
                 & ---  & ---        & ---  \\
\hline
{\em UBVRIK} &(d)& 0.06 & $ 0\pm 0$  & $ 9\pm 11$ 
                 & 0.19 & $ 3\pm 1$  & $22\pm 2$ 
                 & 0.36 & $ 5\pm 1$  & $ 7\pm 1$ 
                 & 0.44 & $14\pm 1$  & $ 2\pm 1$ 
                 & 0.39 & $14\pm 3$  & $ 3\pm 2$ \\
             &(s)& 0.17 & $14\pm 1$  & $14\pm 2$ 
                 & 0.25 & $ 7\pm 1$  & $ 4\pm 1$ 
                 & 0.43 & $11\pm 2$  & $ 2\pm 1$ 
                 & 0.54 & $34\pm 6$  & $25\pm 7$ 
                 & ---  & ---        & ---  \\
\hline
{\em BVRIJK} &(d)& 0.06 & $ 0\pm 0$  & $ 7\pm 6$ 
                 & 0.17 & $ 3\pm 1$  & $32\pm 3$ 
                 & 0.39 & $ 5\pm 2$  & $ 9\pm 1$ 
                 & 0.43 & $25\pm 3$  & $ 4\pm 1$ 
                 & 0.40 & $29\pm 2$  & $ 3\pm 3$ \\
             &(s)& 0.19 & $17\pm 2$  & $ 6\pm 2$ 
                 & 0.24 & $ 7\pm 1$  & $ 2\pm 1$ 
                 & 0.39 & $ 8\pm 1$  & $ 4\pm 1$ 
                 & 0.37 & $30\pm 6$  & $58\pm 8$ 
                 & ---  & ---        & ---  \\
\hline
{\em UBVRIJK}&(d)& 0.06 & $ 0\pm 0$  & $ 5\pm 5$ 
                 & 0.17 & $ 3\pm 1$  & $23\pm 4$ 
                 & 0.35 & $ 3\pm 1$  & $ 8\pm 1$ 
                 & 0.41 & $13\pm 3$  & $ 4\pm 1$ 
                 & 0.38 & $26\pm 4$  & $ 2\pm 2$ \\
$S/N=1$      &(s)& 0.17 & $13\pm 2$  & $12\pm 3$ 
                 & 0.24 & $ 6\pm 1$  & $ 3\pm 1$ 
                 & 0.40 & $ 6\pm 2$  & $ 3\pm 1$ 
                 & 0.43 & $29\pm 7$  & $34\pm 9$ 
                 & ---  & ---        & ---  \\
$S/N=5$
             &(s)& 0.09 & $ 2\pm 1$  & $ 6\pm 2$ 
                 & 0.15 & $ 1\pm 1$  & $ 3\pm 1$ 
                 & 0.40 & $ 8\pm 2$  & $ 2\pm 1$ 
                 & ---  & ---        & ---  
                 & ---  & ---        & ---  \\
$S/N=10$
             &(s)& 0.06 & $ 0\pm 0$  & $ 2\pm 2$ 
                 & 0.11 & $ 0\pm 0$  & $ 2\pm 1$ 
                 & 0.27 & $ 7\pm 10$ & $ 5\pm 4$ 
                 & ---  & ---        & ---  
                 & ---  & ---        & ---  \\
\hline
{\em UBVRI}  &(d)& 0.05 & $ 0\pm 0$  & $ 5\pm 6$ 
                 & 0.17 & $ 3\pm 1$  & $24\pm 2$ 
                 & 0.35 & $ 3\pm 1$  & $ 8\pm 1$ 
                 & 0.39 & $13\pm 2$  & $ 6\pm 1$ 
                 & 0.38 & $27\pm 3$  & $ 2\pm 1$ \\
{\em JHK}    &(s)& 0.17 & $13\pm 2$  & $ 9\pm 2$ 
                 & 0.24 & $ 5\pm 1$  & $ 3\pm 1$ 
                 & 0.38 & $ 5\pm 1$  & $ 4\pm 1$ 
                 & 0.36 & $23\pm 4$  & $41\pm 8$ 
                 & ---  & ---        & ---  \\
\hline
\end{tabular}
\caption{The dispersion $\sigma_z$ and the percentage of catastrophic
and spurious objects, $l$\% and $g$\%, with errors, in five
redshift bins, computed from 10 realizations of simulated
catalogues with a redshift distribution derived from a PLE model. 
(s) and (d) refer to shallow and deep surveys respectively.  The data
are replaced by a dash when there are no enough data to compute the
statistics.  For the three examples in Figures~\ref{figreal1},
\ref{figreal2} and ~\ref{figreal3} we also present the quantities
mentioned above as a function of the limiting signal to noise ratio
considered for the detection.}
\label{newgain}
\end{center}
\end{table*}

In the second case, the aim is to reproduce the observational
conditions reached when using 8\,m telescopes and a wide field
detector. In particular, we consider the case of a survey in a $\sim
60$\,arcmin$^2$ field, observed with all the filter sets considered in
Section~\ref{simul}.  The adopted limiting magnitudes are shallower
and conservative with respect to the values in the previous
simulations. They are shown in the column $m_{\rm lim}$(s) of
Table~\ref{tabmlim}.  The percentages of the different spectral types
for catalogues with these limiting magnitudes, using the same
detection criteria, are similar to the previous ones, being $\sim 12,
54, 26, 8$\% for E, Sb, Im, and Im(${\rm age} = 0.1$\,Gyr)
respectively.

Results for $\sigma_z$, $l$\% and $g$\% are presented in
Table~\ref{newgain} and marked by a (s).  For the filter sets {\em
UBVRI} and {\em UBVRIJK} we repeated the same procedure adopted for
the HDF-N simulated catalogue, to build two subcatalogues with higher
$S/N$ thresholds.  Figure~\ref{figreal2} presents the results
obtained with the five optical bands only, whereas
Figure~\ref{figreal3} displays the equivalent results with the
additional photometry in two near infrared filters.
Figures~\ref{figreal2} and \ref{figreal3} show the associated input
and recovered redshift distributions.

The peak of the redshift distribution in this case is at a lower
redshift compared to the HDF simulation. Wide-field surveys allow to
obtain a better sampling of the bright end of the luminosity function
with respect to HDF-like surveys, the later being more suited to
explore the faint luminosity regime.
The value of $g$\% and $l$\% change significantly when considering the
same set of filters, but a different kind of survey.  On the contrary,
$\sigma_z$ remains similar.

An interesting feature is the opposite trend displayed by deep
pencil-beam compared to shallow wide-field surveys with respect to the
low and high redshift regimes for a given filter set.  At low
redshifts, the value of $g$\% is larger for the deep pencil-beam
survey (type (d) in the Table~\ref{newgain}) than for the
shallow (s) wide-field one.  Conversely, the accuracy of deep surveys
overcome that of the shallow ones at high redshifts. In this context,
the separation between low and high redshift regimes is marked by the
$z=1$ -- $2$ bin.

This behaviour could be easily explained when we consider the
different characteristics of the catalogues produced in the two cases.
The deep survey catalogue contains few low redshift galaxies, and most
of them derive from the faint tail of the luminosity function.
These faint galaxies are much more abundant than the bright ones, than
they are present in the catalogue even though the volume covered at
low redshift by this survey is small.  The photometric errors for
these intrinsically faint objects are rather large, thus causing a
poor estimate of $z_{\rm phot}$.
On the contrary, the wide-field survey contains a large quantity of
bright galaxies at low redshift, wich have sufficiently small
photometric errors to obtain accurate $z_{\rm phot}$s. The faintest
objects are lost in this case because of the shallow detection
limits. The majority of galaxies in the shallow wide survey lies in
the low redshift bins, around the peak of $N(z)$.  When we consider
the population of galaxies beyond the peak of the redshift
distribution, the photometric errors in the shallow survey become
important and an increasing fraction of objects is non detected in
various filters.  These problems hamper a robust determination of
$z_{\rm phot}$.  On the contrary, the pencil beam survey take
advantage of its depth, allowing to compute $z_{\rm phot}$ at higher
redshifts.

On the basis of these results, we caution that the kind of analysis
presented here is strongly advised when a photometric survey is
undertaken in view of computing $z_{\rm phot}$s. In particular, the
filter configuration and the photometric depth to reach in each filter
have to be determined accurately in advance, in order to optimize the
survey and to study the feasibility of the project.

\section{Discussion}
\label{discuss}

Making use of the Bayesian technique, Ben\'{\i}tez (2000) demonstrated
that the dispersion of $z_{\rm phot}$ can be significantly improved.
Despite of this result, we decide not to introduce this possibility in
our code, at least for general purposes. The reason for this is that
we want to prevent spurious effects in particular studies. As an
example, when the luminosity function is imposed, the study of the
galaxy population is constrained and it becomes impossible to obtain
independent information on the properties of objects, thus limiting
the possible applications.
However, this method can be regarded with interest when the purpose is
addressed to some specific application or when one is dealing with
poor data, in such a way that the introduction of hints permits to obtain
useful results. 
Alternatively, the photometric redshift estimate can be safely
improved introducing the Bayesian inference when prior information is
not related to photometric properties of sources. Examples of such priors
that could be combined with the $z_{\rm phot}$ technique are 
the morphology or the clues inferred from gravitational lensing
modeling.

   One of the main issues for $z_{\rm phot}$ is the optimization of
the visible versus near-IR bands for spectroscopic surveys. The aim is
to produce a criterion based in $z_{\rm phot}$ to discriminate between
objects showing strong spectral features in the optical and in the
near-IR. To perform this test, both the redshift and the SED
characteristics have to be estimated for each object. The $z_{\rm
phot}$ and the SED are obtained by means of {\it hyperz\/}, together
with the best fit parameters ($A_V$, spectral type, metallicity and
age). The relevant information shall be the redshift and the rough SED
type, i.e. ``blue'' or ``red'' continuum at the given $z$.
We have shown that only limited information could be obtained 
on the parameter space from broad-band photometry alone. 
This situation will change with the future cryogenic imaging
spectrophotometers, as presented in a recent paper by Mazin \& Brunner
(2000), because such devices will be able to gain in spectral
resolution while spanning a large wavelength domain.

   Another important issue for $z_{\rm phot}$ is the improvement on
the cluster detection in wide-field photometric surveys. Including
such a technique in an automated identification algorithm, whatever
this algorithm is, allows to improve significantly the detection
levels. The main idea is that the contrast between the cluster and the
foreground and background population is the leading factor. When
introducing a simple detection scheme, similar to the one used by
Cappi et al. (1989), it is easy to quantify this effect (Pell\'o et
al. 1998). In general, the $S/N$ is expected to improve by a factor of
at least $\sim 2$ to $3$ with respect to the pure 2D case, depending
on the cluster redshift and richness, the set of filters used and the
depth of the survey. When considering more elaborated cluster-finding
algorithms, such as the one produced by Kepner et al. (1999), Olsen et
al. (1999), Scodeggio et al. (1999), Kawasaki et al. (1998) or Deltorn
et al. (2000, in preparation), these results could be regarded as the
relative improvement due to photometric redshifts.
The present version of {\it hyperz\/} is also able to display the
probability of each object to be at a fixed redshift. This is useful
when looking for clusters of galaxies at a given (guessed) redshift.

   The study of clustering properties through the spatial correlation
function of galaxies, using the angular correlation together with the
$z_{\rm phot}$ information is another possible application of $z_{\rm
phot}$, aiming to extend the study of galaxy properties to fainter
limits in magnitude. In this case, the relatively high number of
objects accessible to photometry per redshift bin, suitably defined
according to photometric redshift accuracy, allows to enlarge the
spectroscopic sample towards the faintest magnitudes, and also to
strongly reduce the errors (because the number of objects per redshift
bin strongly increases).
Studies on the evolution of the angular correlation function of
galaxies in the HDF-N applying the photometric redshift technique can
be found in Miralles \& Pell\'o (1998), Connolly et al. (1998),
Roukema et al. (1999), Arnouts et al. (1999), Magliocchetti \& Maddox
(1999), .

The same slicing procedure can be adopted to study the evolution of
the luminosity function and consequently to infer the star formation
history at high redshift from the UV luminosity density, as well as
to analyse the stellar population and the evolutionary properties of
distant galaxies (e.g. Yee et al. 1996, SubbaRao et al. 1996, Gwyn \&
Hartwick 1996, Sawicki et al. 1997, Connolly et al. 1997, Pascarelle
et al. 1998, Giallongo et al. 1998).

Furthermore, the photometric redshift method has been used to
investigate the nature of Extremely Red Objects (EROs) with a
``spectro-photometric'' technique by Cimatti et al. (2000), deducing
clues about the model of galaxy formation.
Another kind of spectroscopic and photometric combination has led to
the identification of very high redshift object, as described by Chen et
al. (1999).

From this not exhaustive list of applications, it is evident that
photometric redshifts are a powerful and promising tool in many areas
of extragalactic research. This method shall not be regarded only as a
``poor person's redshift machine'', but as a fundamental instrument,
since a multitude of faint objects will remain beyond the limits of
spectroscopy for the next years. Even with the diffusion of Multi-Object 
Spectrometers, most of the faint galaxies with measured photometry will
fall beyond the reach of conventional spectroscopy.

\section{Conclusions}
\label{conclu}

We have presented the characteristics and the performances of our
public code {\it hyperz}, available on the web, which make use of the
template SED fitting technique.  We can summarize the main conclusions
as follows:

\begin{enumerate}

\item Simulations of ideal catalogues have shown the main trends of
the accuracy on $z_{\rm phot}$ calculations. In particular, $z_{\rm
phot}$ estimates are improved when the filters set spans a wide
wavelength range, including near-IR and $U$ filters, and when the
photometric errors become small.

\item We have investigated the weight of the different parameters on
the final results, using both a spectroscopic subsample of HDF and
simulations. In particular, the templates, the flux decrement by Lyman
forest, the age of the stellar population, the reddening, the
cosmology, the metallicity, the IMF and the presence of emission lines
have been discussed. According to these results, the $z_{\rm phot}$
preciseness seems to be more sensitive to the photometric accuracy
rather than to the detailed set of parameters. Nevertheless, a subset
of these parameters (reddening, age of the stellar population and
Lyman forest blanketing) has to span a sufficiently wide range of
values to obtain accurate $z_{\rm phot}$s.

\item The robustness of the method has been illustrated through
realistic deep field simulations, aiming to reproduce the redshift
distribution, photometric accuracy and limiting magnitudes encountered
in deep field surveys.

\item We have pointed out some of the manifold applications of the
photometric redshift machinery in present and future projects.

\item We plan to include AGN SEDs in the present scheme of {\it hyperz\/},
as well as stellar templates, in order to automatically classify
objects in a photometric survey through a unique pipeline. This
particular application is presently under development (Hatziminaoglou
et al. 2000).

\end{enumerate}

\acknowledgements

We are grateful to S. Charlot, G. Bruzual, M. Dantel-Fort, G. Mathez, 
E. Hatziminaoglou, J.P. Kneib, Y. Mellier, D. Maccagni, D. Valls-Gabaud
and J.-P. Picat for interesting comments and discussion.
This work has been done within the framework of the VIRMOS 
collaboration.
We would thank L. Moscardini and E. Carretta for useful suggestions 
on the preliminary version of the manuscript. 
Part of this work was supported by the French {\it Centre National de
la Recherche Scientifique}, by the French {\it Programme National de 
Cosmologie} (PNC), and the TMR {\it Lensnet} ERBFMRXCT97-0172
(http://www.ast.cam.ac.uk/IoA/lensnet).
J.-M.M. thanks the JSPS Postdoctoral Fellowship Grant (P98176) for 
financial support.


\begin{thebibliography}{}

\bibitem{} Allen, C.W. 1976, ``Astrophysical Quantities'', University
of London, The Athlone Press, pag. 264

\bibitem{} Arnouts, S., Cristiani, S., Moscardini, L., Matarrese, S.,
Lucchin, F., Fontana, A., Giallongo, E. 1999, MNRAS 310, 540

\bibitem{} Avni, Y. 1976, ApJ 210, 642

\bibitem{} Baum, W.A. 1963, IAU Symposium n. 15, Macmillan Press, New
York, p.390

\bibitem{} Ben\'{\i}tez, N. 2000, ApJ 536, 571

\bibitem{} Bertin, E., Arnouts, S. 1996, A\&ASS 117, 393

\bibitem{} Biretta, J.A., et al. 1996, WFPC2 Instrument Handbook, 
Version 4.0, STScI publications

\bibitem{} Bouchet, P., Lequeux, J., Maurice, E., Pr\'evot, L.,
Pr\'evot-Burnichon, M. L. 1985, A\&A 149, 330

\bibitem{} Brunner, R.J., Connolly, A. J., Szalay, A.S., Bershady,
M.A.  1997, ApJ 482, 21

\bibitem{} Bruzual, G., Charlot, S. 1993, ApJ 405, 538

\bibitem{} Calzetti, D., Armus, L., Bohlin, R.C., Kinney, A.L.,
Koornneef J., Storchi-Bergmann T. 2000, ApJ 533, 682

\bibitem{} Cappi, A., Chincarini, G., Conconi, P., Vettolani, G., 1989
A\&A 223, 1.

\bibitem{} Chen, H.-W., Lanzetta, K.M., Pascarelle, S. 1999, 
Nature 398, 586

\bibitem{} Cimatti, A., Daddi, E., di Serego Alighieri, S., Pozzetti, L.,
Mannucci, F., Renzini, A., Oliva, E., Zamorani, G., Andreani, P.,
R\"ottgering, H.J.A., 2000, A\&A 532, L45

\bibitem{} Cohen, J.G., Cowie, L.L., Hogg, D.W., Songaila, A., Blandford,
 R., Hu, E.M., Shopbell, P. 1996, ApJ 471, L5

\bibitem{} Coleman, D.G., Wu, C.C., Weedman, D.W. 1980, ApJS 43, 393

\bibitem{} Connolly, A.J., Csabai, I., Szalay, A.S., Koo, D.C., Kron, R.G., 
Munn, J.A. 1995, AJ 110, 2655

\bibitem{} Connolly, A.J., Szalay, A.S., Dickinson, M., SubbaRao, M.U., 
Brunner, R.J. 1997, ApJ 486, L11

\bibitem{} Connolly, A.J., Szalay, A.S., Brunner, R.J. 1998, ApJ 499, L125

\bibitem{} Couch, W.J., Ellis, R.S., Godwin, J., Carter, D. 1983, 
MNRAS 205, 1287

\bibitem{} Cowie, L.L.  1997, \\
{\tt http://www.ifa.hawaii.edu/cowie/tts/tts.html}

\bibitem{} Cowie, L.L., Hu, E.M., Songaila, A. 1995, Nature 377, 603

\bibitem{} Dickinson, M., et al. 2000, in preparation

\bibitem{} Fern\'andez-Soto, A., Lanzetta, K.M., Yahil, A. 1999, ApJ
513, 34

\bibitem{} Fitzpatrick, E.L. 1986, AJ 92, 1068

\bibitem{} Furusawa, H., Shimasaku, K., Doi, M., Okamura, S. 2000, 
ApJ 534, 624

\bibitem{} Giallongo, E., Cristiani, S. 1990, MNRAS 247, 696

\bibitem{} Giallongo, E., D'Odorico,, S., Fontana, A., Cristiani, S., 
Egami, E., Hu, E., McMahon, R.G. 1998, AJ 115, 2169

\bibitem{} Glazebrook, K., Ellis, R.S., Colless, M., Broadhurst, T.,
Allington-Smith, J., Tanvir, N. 1995, MNRAS 273, 157

\bibitem{} Guzm\'an, R., Gallego, J., Koo, D.C., Phillips, A.C.,
Lowenthal, J.D., Faber, S.M., Illingworth, G.D., Vogt, N.P. 1997,
ApJ 489, 559

\bibitem{} Gwyn, S.D.J., Hartwick F.D.A. 1996, ApJ 468, L77

\bibitem{} Hatziminaoglou, E., Mathez, G., Pell\'o, R. 2000,
A\&A 359, 9

\bibitem{} Hogg, D. W., Cohen, J.G., Blandford, R., et al. 1998, AJ
115, 1418

\bibitem{} Kawasaki, W., Shimasaku, K., Doi, M., Okamura, S. 1998, A\&AS
130, 567

\bibitem{} Kepner, J., Fan, X., Bahcall, N., Gunn, J., Lupton, R. 1999,
ApJ 517, 78

\bibitem{} Koo, D.C. 1985, AJ 90, 418

\bibitem{} Lanzetta, K.M., Yahil, A., Fern\'andez-Soto, A. 1996, 
Nature 381, 759

\bibitem{} Lowenthal, J.D., Koo, D.C., Guzm\'an, R., et al. 1997, ApJ
 481, 673 

\bibitem{} Madau, P. 1995, ApJ 441, 18

\bibitem{} Magliocchetti, M., Maddox, S.J. 1999, MNRAS 306, 988

\bibitem{} Mazin, B.A., Brunner R.J. 2000, AJ accepted
[astro-ph/0007420]

\bibitem{} Miller, G.E., Scalo, J.M. 1979, ApJS 41, 513

\bibitem{} Miralles, J.M. 1998, PhD. thesis {\it Universit\'e Paul Sabatier}

\bibitem{} Miralles, J.M., Pell\'o, R., ApJ submitted [astro-ph/9801062]

\bibitem{} Mobasher, B., Mazzei, P. 1999, Proceedings of
``Photometric Redshifts and High Redshift Galaxies'', Weymann et
al. Eds., Astr. Soc. of Pac. Conf. Series 191 [astro-ph/9907210]

\bibitem{} Mobasher, B., Rowan-Robinson, M., Georgakakis, A., Eaton, N. 
1996, MNRAS 282, L7

\bibitem{} Olsen, L.F., Scodeggio, M., da Costa, L., et al.
%Benoist, C., Bertin, E., Deul, E., Erben, T., Guarnieri, M. D., 
%Hook, R., Nonino, M., Prandoni, I., Slijkhuis, R., Wicenec, A., 
%Wichmann, R., 
1999, A\&A 345, 681

\bibitem{} Pascarelle, S.M., Lanzetta, K.M., Fern\'andez-Soto, A. 
1998, ApJ 508,L1

\bibitem{} Pell\'o, R., Kneib, J.-P., Bolzonella, M., Miralles, J.-M. 
1999, Proceedings of the ``Photometric Redshifts and High
Redshift Galaxies'', Weymann et al. Eds., Astr. Soc. of
Pac. Conf. Series 191, p. 241.  [astro-ph/9907054].

\bibitem{} Pell\'o, R., Leborgne, J.F., Miralles, J.M., Bruzual, G.  1998,
{\it Detection of distant clusters of galaxies using photometric redshifts} 
Proceedings of the 14th IAP Meeting ``Wide Field Surveys in Cosmology'', 
Colombi S. et al. Eds., p. 410.

\bibitem{} Pozzetti, L., Bruzual, G., Zamorani, G. 1996, MNRAS 281, 953

\bibitem{} Pozzetti, L., Madau, P., Zamorani, G., Ferguson, H.,
Bruzual, G.  1998, MNRAS 298, 1133

\bibitem{} Pr\'evot, M.L., Lequeux, J., Pr\'evot, L., Maurice, E.,
Rocca-Volmerange, B. 1984, A\&A 132, 389

\bibitem{} Roukema, B.F., Valls-Gabaud, D., Mobasher, B., Bajtlik, S. 1999,
MNRAS 305, 151

\bibitem{} Salpeter, E.E. 1955, ApJ 121, 161

\bibitem{} Sawicki, M.J., Lin, H., Yee, H.K.C. 1997, AJ 113, 1

\bibitem{} Scalo, J.M. 1986, Fundamentals of Cosmic Physics 11,  p. 1-278

\bibitem{} Scodeggio, M., Olsen, L.F., da Costa, L., Slijkhuis, R., et al. 
%Benoist, C., Deul, E., Erben, T., Hook, R., Nonino, M., 
%Wicenec, A., Zaggia, S., 
1999, A\&AS 137, 83

\bibitem{} Seaton, M.J. 1979, MNRAS 187, 73

\bibitem{} Steidel, C.C., Adelberger, K.L., Giavalisco, M., Dickinson,
M., Pettini, M. 1999, ApJ 519, 1

\bibitem{} Steidel, C.C., Giavalisco, M., Pettini, M., Dickinson, M.,
Adelberger, K.L. 1996, ApJ 462, L17

\bibitem{} SubbaRao, M.U., Connolly, A.J., Szalay, A.S., Koo, D.C. 
1996, AJ 112, 929

\bibitem{} Terlevich, R., Melnick, J., Masegosa, J., Moles, M.,
Copetti, M.V.F. 1991, A\& AS 91, 285

\bibitem{} Trager S.C., Faber S.M., Dressler A., Oemler A. 1997,
ApJ 485, 92.

\bibitem{} Wang, Y., Bahcall, N., Turner, E.L. 1998, AJ 116, 2081 

\bibitem{} Williams, R.E., Blacker, B., Dickinson, M., et al. 1996, AJ
112, 1335

\bibitem{} Yee, H.K.C., Ellingson, E., Bechtold J., Carlberg, R.G.,
Cuillandre J.-C. 1996, AJ 111, 1783

\bibitem{} Zepf, S.E., Moustakas, L.A., Davis, M.  1997, ApJ 474, L1

\end{thebibliography}
\end{document}